\documentclass{aa}

\usepackage{graphicx}
\usepackage{txfonts}
\usepackage{amsmath}
\usepackage{amssymb}
\usepackage{color}
\usepackage{natbib}
\usepackage{hyperref}

\makeatletter
\renewcommand*\aa@pageof{, page \thepage{} of \pageref*{LastPage}}
\makeatother

\newcommand{\Msun}{$\mathrm{M}_{\odot}$}
\newcommand{\Rsun}{$\mathrm{R}_{\odot}$}

\newcommand*\mean[1]{\overline{#1}}

\newcommand{\mlt}{SSM-MLT}
\newcommand{\ssmoneeq}{SSM-1KM}
\newcommand{\ssmthreeeq}{SSM-3KM}
\newcommand{\oneeq}{1KM}
\newcommand{\threeeq}{3KM}

\defcitealias{Magg2022}{M22}
\defcitealias{Asplund2009}{A09}
\defcitealias{Grevesse1998}{GS98}

\begin{document}

   \title{Testing the 3-equation Kuhfuss Convection Model using the Sun}

   \author{T. A. M. Braun
          \inst{1, 2}\thanks{braun@mpa-garching.mpg.de}
          \and
          F. Ahlborn\inst{3}
        \and
        F. Kupka\inst{4, 5, 6}
        \and
        A. Weiss\inst{1}
          }

   \institute{Max-Planck-Institut f{\"u}r Astrophysik, Karl-Schwarzschild-Straße 1, 85741 Garching, Germany
    \and
    Ludwig-Maximillians-Universit\"at M\"unchen, Geschwister-Scholl-Platz 1, 80539 Munich, Germany
    \and
    Heidelberger Institut für Theoretische Studien, Schloss-Wolfsbrunnenweg 35, 69118 Heidelberg, Germany
    \and
    Faculty of Comp. Sci. \& Appl. Math., Univ. of Applied Sciences, Technikum Wien, H\"ochst\"adtplatz 6, A-1200 Wien, Austria
    \and
    Wolfgang-Pauli-Institute c/o Faculty of Mathematics, University of Vienna, Oskar-Morgenstern-Platz 1, A-1090 Wien, Austria
    \and
    Fakult\"at f\"ur Mathematik, Universit\"at Wien, Oskar-Morgenstern-Platz 1, A-1090 Wien, Austria
    \\
             }

   \date{}

  \abstract
   {Simplified, one-dimensional models are necessary to model convection in the context of stellar evolution. By including the non-local effects of convection, turbulent convection models describe convection in a more physical way compared to mixing length theory, which is typically used in one-dimensional stellar evolution models. We recently showed that the 1-equation Kuhfuss turbulent convection model is not sufficient to model the solar convective envelope satisfactorily.}
   {Using the Sun as a benchmark, we test the physically more complete 3-equation Kuhfuss turbulent convection model.}
   {We calculate a solar calibrated model with the 3-equation Kuhfuss turbulent convection model using the one-dimensional stellar evolution code GARSTEC. We compare the predicted interior structure of the model with helioseismic measurements of the Sun. Furthermore, we investigate how the free parameters and the closure relations of the 3-equation model influence the results.}
   {We find that, with the 3-equation model, the temperature gradient at the inner boundary of the convective envelope is modelled more realistically compared to the mixing length theory or the 1-equation model. This also improves the agreement for the sound speed profile between the model and the Sun, and reduces the asteroseismic surface effect. However, close to the surface, the 3-equation model results in a layer having an unphysical, negative temperature gradient. This layer is connected to the closure relations used in the 3-equation model.}
   {Our results demonstrate the capabilities of turbulent convection models, and can serve as a next step towards an improved and more realistic modelling of convection in stellar evolution codes.}


   \maketitle

\section{Introduction}\label{sec:intro}

Convection is one of the main energy transport mechanisms in stars, and very efficient in mixing chemical elements. Therefore, it is a crucial ingredient in stellar evolution.
Convection is an inherently three-dimensional (3D) process, and 3D simulations are important to study the detailed dynamics in a convective zone \citep[see ][and references therein]{Kapyla2023, Lecoanet2023}. 
However, when studying stellar evolution, one-dimensional (1D) stellar evolution codes are needed. This is because modelling convection in 3D is computationally expensive and often not feasible when timescales of stellar evolution are considered \citep{Kupka2017, Lecoanet2023}. That is due to the several orders of magnitude between the timescale important for convection and that important for stellar evolution.
Thus, 1D stellar evolution codes will not become obsolete and improving their input physics will remain important \citep{Kupka2017}.

The modelling of convection in 1D stellar evolution codes goes back to Ludwig Prandtl who, in analogy to the mean free path in diffusion processes of gas, introduced the mixing length as a length scale to describe the effects of turbulence \citep{Prandtl1925}. Ludwig Biermann applied this theory to stars \citep{Biermann1932b, Biermann1948}. This model, mainly in the formulations developed by \citet{Bohm-Vitense1958} and \citet{Cox1968}, is known as Mixing Length Theory (MLT) and is widely used in 1D stellar evolution codes \citep[e.g., MESA, CESAM, GARSTEC, ][respectively; for a review see \citealt{Joyce2023}]{Jermyn2023, Morel2008, Weiss2008}. It is the most frequently used version among all local convection theories.
However, already \citet{Prandtl1925} warned that MLT is merely a crude approximation. This assessment was soon confirmed by disagreements between stellar models and observations.  
For example, the predicted temperature profile in the envelope of a solar model does not match the solar temperature profile. This leads to disagreements between the theoretical and observed frequencies, the so-called asteroseismic surface effect \citep{Christensen-Dalsgaard1996}.
Another example is the convectively mixed core sizes of intermediate-mass main-sequence stars. MLT predicts these to be too small, resulting in too short main-sequence lifetimes \citep[e.g.,][]{Napiwotzki1991, Chiosi1992, Zhang2012, Claret2016, Tkachenko2020}.

Disagreements between stellar models and observations, such as convective core sizes being too small, are influenced by the local description of convection in MLT. In a local convection model, the decision as to whether a layer is convective or not, is solely based on the state of the fluid at this specific, local layer. In the framework of MLT, if chemical gradients are neglected, this implies that all convective motion comes to a halt at the Schwarzschild boundary. This is the radius where the radiative and adiabatic temperature gradient are equal ($\nabla_\mathrm{rad}=\nabla_\mathrm{ad}$). As a consequence, in MLT convective motions immediately cease at the same point where the buoyancy force, which is the main driving force of convection, disappears.
However, due to the inertia of the fluid parcels, it is expected that they penetrate, or ``overshoot'' into formally stable layers. The fluid parcels are decelerated until they reach a velocity of zero, thereby transporting energy and elements beyond unstable, convective regions. To include this non-local process in a convection model, it needs to be considered that layers through which a fluid parcel travelled at earlier times also influence the characteristics of the fluid parcel. All such non-local processes are generally called convective boundary mixing \citep[CBM, see the review by ][]{Anders2023}. There are subcategories based on the effect CBM has on the mixing of elements and the temperature stratification. Overshooting refers to CBM which only extends the chemically mixed region by a certain amount and does not change the temperature stratification. If, in addition to the extended chemical mixing, the temperature stratification in the CBM layer is altered to be close-to-adiabatic, it is called ``adiabatic overshooting'', or ``convective penetration'' \citep{Zahn1991}. Convective boundary mixing is observed in 3D simulations \citep[e.g.,][]{Hurlburt1986, Freytag1996, Kapyla2019} and is needed to bring models and observations into agreement \citep[e.g.,][]{Napiwotzki1991, Alongi1991, Chiosi1992, Demarque1994}. MLT does not account for CBM, and additional, mostly parametrized overshooting descriptions were introduced {\em ad hoc} to bring models and observations into agreement \citep[e.g.,][]{Shaviv1973, Maeder1975, Freytag1996, Anders2023}.

More physically complete turbulent convection models (TCMs) are available, for example \citet{Xiong1997, Kuhfuss1986, Kuhfuss1987, Canuto1998, Deng2006, Li2007}. These convection models have the advantage of including the non-local effects already from the start. However, due to the turbulent, and thus, highly non-linear nature of convection, they are more difficult to include in a stellar evolution code. So far, only a few studies within the framework of stellar evolution codes have been done \citep[e.g., ][]{Zhang2012b, Zhang2012c, Zhang2012, Ahlborn2022, Braun2024, Deka2025}.

Kuhfuss developed two models, which fall into these more advanced frameworks \citep{Kuhfuss1986, Kuhfuss1987}: the 1- and 3-equation Kuhfuss convection model (\oneeq\ and \threeeq). Both models are implemented in the GARching STellar Evolution Code \citep[GARSTEC, ][]{Weiss2008}, and were improved by \citet{Wuchterl1998}, \citet{Flaskamp2003}, \citet{Kupka2022}, and \citet{Ahlborn2022}. Both models include non-local effects of convection. The \threeeq\ is physically more complete in comparison to the \oneeq, since it accounts for countergradient fluxes (i.e., positive enthalpy fluxes in locally stably stratified regions; see Sect.~\ref{sec:discussion} for more details on the differences between \oneeq\ and \threeeq). 

A new convection model first needs to be verified and tested against benchmarks. So far, both Kuhfuss convection models have been successfully tested against the size of convective cores of intermediate-mass main-sequence stars \citep{Ahlborn2022}. 
Applying the \oneeq\ to Cepheid stars in binary systems confirmed that it can reproduce the observations as well as models applying MLT with ad hoc overshooting \citep{Deka2025}.
In a previous paper \citep{Braun2024}, we applied the \oneeq\ in a standard solar model (SSM). It resulted in an adiabatic overshooting layer at the bottom of the convective envelope, leading to a strong disagreement with helioseismic inferences of the solar structure.

Building on these results, we again use the Sun to test the more physically complete \threeeq. 
We find that the \threeeq\ improves the profile of the temperature gradient in the CBM region. The gradual change of the temperature gradient also significantly improves the agreement of the solar model's sound speed profile with the sound speed profile of the Sun.

The paper is structured as follows: An overview over \threeeq\ is provided in Sect.~\ref{sec:kuhfuss}. In Sect.~\ref{sec:obsandmodels}, we shortly discuss the available observables of the Sun and the stellar models we calculated, which are then compared in Sect.~\ref{sec:results}. We close this paper with a discussion (Sect.~\ref{sec:discussion}) and a summary and conclusion (Sect.~\ref{sec:conclusion+summary}). We briefly comment on the effect of an alternative solar composition in Appendix~\ref{ap:asplund-abundances}.

\section{The 3-equation Kuhfuss Model}\label{sec:kuhfuss}

To derive the 3-equation Kuhfuss model (\threeeq), the Reynolds splitting \citep{Reynolds1895} was applied to the Navier-Stokes equation and the energy conservation equation. That means, hydrodynamic quantities $a$ were split into a spherically averaged component $\mean{a}$ and a fluctuating component $a'$.
The \threeeq\ \citep{Kuhfuss1987} consists of equations for the turbulent kinetic energy (TKE) $\omega=\frac{1}{2}\mean{\boldsymbol{u}'^2}$, the second order entropy fluctuations $\Phi=\frac{1}{2}\mean{s'^2}$, and the correlation of velocity and entropy fluctuations $\Pi=\mean{s'u'_r}$, which is connected to the convective flux $F_\mathrm{conv}$ by 
\begin{equation}
    F_\mathrm{conv} = \rho T \Pi \, ,
\end{equation}
with the density $\rho$ and the temperature $T$. The entropy and velocity fluctuations are denoted by $s'$ and $\boldsymbol{u}'$, respectively. $u'_r$ is the radial component of the velocity fluctuations.
The dynamical equations for $\omega$, $\Pi$, and $\Phi$ are given as follows:

\begin{align}
    \frac{\partial \omega}{\partial t} &= \frac{\nabla_\mathrm{ad} T}{H_p} \Pi  &- \frac{C_D}{\Lambda} \omega^{3/2}     &                                 &- \mathcal{F}_\omega  \, ,  \label{eq:tke}\\
    \frac{\partial \Pi}{\partial t}    &= \frac{2\nabla_\mathrm{ad}T}{H_p}\Phi  &+\frac{2c_p}{3H_p}(\nabla-\nabla_\mathrm{ad})\omega &- \frac{1}{\tau_\mathrm{rad}}\Pi &- \mathcal{F}_\Pi  \, ,  \label{eq:pi}\\
    \frac{\partial \Phi}{\partial t}   &=                                       &\frac{c_p}{H_p}(\nabla-\nabla_\mathrm{ad})\Pi      &- \frac{2}{\tau_\mathrm{rad}}\Phi &- \mathcal{F}_\Phi \, . \label{eq:phi}
\end{align}
The bar for mean quantities was omitted to enhance readability. These equations can also describe the time dependency of convection. For this work, we concentrate on studying the main-sequence evolution, where the timescale of stellar evolution is long enough to ensure that convection can reach a steady state. Thus, we neglect the time dependency and set $\frac{\partial a}{\partial t}=0$ ($a\in\{\omega ,\Pi, \Phi\}$). In the following, we discuss the individual terms constituting the equations of the \threeeq\ in more detail. For a more detailed derivation of the model, we refer the reader to Appendix A by \citet{Kupka2022}, and the original work by \citet{Kuhfuss1987}. Further details about the implementation of the \threeeq\ can be found in the work by \citet{Flaskamp2003}, \citet{Kupka2022}, and \citet{Ahlborn2022}.

The driving by buoyancy force, caused by the pressure gradient, is described by the first terms of Eqs.~\eqref{eq:tke} and \eqref{eq:pi} with the pressure scale height $H_p$, and the adiabatic temperature gradient $\nabla_\mathrm{ad}$.

The second term in Eq.~\eqref{eq:tke} describes the dissipation of energy, where the dissipation law from \citet{Kolmogorov1941, Kolmogorov1962} was used. The parameter $C_D$ is a free parameter.
Following \citet{Wuchterl1995}, the parameter $\Lambda$ is calculated by 
\begin{align}\label{eq:hp}
    \frac{1}{\Lambda} &= \frac{1}{\alpha H_p} + \frac{1}{\beta r} \, ,
\end{align}
where $r$ is the radius, $\alpha$ is a free parameter which is typically of order unity, and $\beta$ is given by the modification derived by \citet{Kupka2022} to include the dissipation by buoyancy waves:
\begin{equation}\label{eq:buoyancywaves}
    \beta = \begin{cases}
    1 & \text{if } \nabla-\nabla_\mathrm{ad} > 0 \\
    (1 + c_4 \Lambda \omega^{-1/2} \Tilde{N})^{-1} &\text{otherwise.}
    \end{cases}
\end{equation}
The Brunt-Väisälä frequency $\Tilde{N}$ is given by 
\begin{equation}
    \Tilde{N}^2 = \frac{g^2 \rho}{p}(\nabla_\mathrm{ad}-\nabla+\nabla_\mu)\, ,
\end{equation}
using the pressure $p$, the gravitational acceleration $g$ and the dimensionless mean molecular weight gradient $\nabla_\mu$. $\nabla$ denotes the dimensionless temperature gradient. The free parameter $c_4$ follows from $c_4=c_3/(c_2 c_{\epsilon})$, using model parameters $c_2=1.92$ and $c_3=0.3$ from \citet{Canuto1998}, and the dissipation rate $c_\epsilon=C_D$, in the case of the Kuhfuss model \citep{Ahlborn2022}.

The negative gradient of the specific entropy ($\partial s/\partial r$) acts as a driving term for $\Pi$ and $\Phi$. It is given as
\begin{equation}
    \frac{\partial s}{\partial r} = - \frac{c_p}{H_p}(\nabla-\nabla_\mathrm{ad}) \, ,
\end{equation}
with the specific heat capacity $c_p$. The effects caused by the entropy gradient are described by the second term in Eq.~\eqref{eq:pi} and the first term in Eq.~\eqref{eq:phi}.
The radiative losses (third term in Eq.~~\eqref{eq:pi}, second term in Eq.~\eqref{eq:phi}) are modelled by a typical timescale for radiative cooling of a convective element $\tau_\mathrm{rad}$:
\begin{equation}
    \tau_\mathrm{rad} = \frac{c_p \kappa \rho^2 \Lambda^2}{4 \sigma T^3 \gamma^2} \, .
\end{equation}
This approximation is based on \citet{Vitense1953} and \cite{Bohm-Vitense1958}. The opacity is denoted as $\kappa$, the Stefan-Boltzmann constant as $\sigma$, and $\gamma$ is a free parameter.

The third order moments (TOMs), denoted as $\mathcal{F}_a$ in Eqs.~\eqref{eq:tke} to \eqref{eq:phi}, are modelled by a diffusion approximation
\begin{equation}\label{eq:non-local-closure}
    \mathcal{F}_a = \mean{(\boldsymbol{u}'\nabla)a} = - \frac{1}{\rho} \boldsymbol{\nabla} \cdot (\alpha_a \Lambda \rho \omega^{1/2} \boldsymbol{\nabla} a) \, ,
\end{equation}
where $a \in \{\omega, \Pi, \Phi\}$, and $\alpha_\omega, \alpha_\Pi, \alpha_\Phi$ are free parameters. These equations introduce the non-local effects into the convection model.

For the equations for $\Pi$ and $\Phi$ (Eq.~\eqref{eq:pi} and \eqref{eq:phi}), \citet{Kuhfuss1987} also derived a local closure relation, which is
\begin{equation}\label{eq:local-closure}
    \mathcal{F}_a = \mean{(\boldsymbol{u}'\nabla)a} = \frac{\alpha_a}{\Lambda} \omega^{1/2} a \, ,
\end{equation}
with $a \in \{\Pi, \Phi\}$.

The temperature gradient is calculated with
\begin{equation}
    \nabla = \nabla_\mathrm{rad}-\frac{H_p \rho}{k_\mathrm{rad}} \Pi \, ,
\end{equation}
where the radiative diffusivity $k_\mathrm{rad}$ is given as
\begin{equation}
    k_\mathrm{rad} = \frac{16 \sigma T^3}{3 \kappa \rho} \, .
\end{equation}

Similar to MLT, mixing in convective regions is modelled with a diffusion process. The convective velocity $v_c$ is obtained from the TKE by assuming full isotropy $v_c = \sqrt{2/3 \omega}$, and the diffusion coefficient $D$ is calculated by 
\begin{equation}
    D = \frac{1}{2} \Lambda \sqrt{\frac{2}{3}\omega} \, ,
\end{equation}

\citep{Ahlborn2022}. In summary, the \threeeq\ has 7 free parameters: The parameters $\alpha_\omega$, $\alpha_\Pi$, and $\alpha_\Phi$ are introduced by the modelling of the TOMs, $\gamma$ influences the timescale of radiative cooling of convective elements, and three more parameters are in the dissipation term of the TKE: $C_D$, $\alpha$, and $c_4$. The parameters $\alpha$, and $c_4$ also appear in the closure relations for the TOMs. We stress that $\alpha$ does not have the same effect as $\alpha_\mathrm{MLT}$ in MLT, due to the calculation of the convective flux from an additional differential equation.

\citet{Kuhfuss1987} calibrated the parameters $C_D$, $\gamma$, $\alpha_\Pi$, and $\alpha_\Phi$ in the local case against MLT. 
This led to $C_D=8/3\sqrt{2/3}$, $\gamma=2\sqrt{3}$, $\alpha_\Pi=6\sqrt{2/3}$, and $\alpha_\Phi=4\sqrt{2/3}$. The overshooting parameter $\alpha_\omega=0.25$ was approximated using the \oneeq\ and considering a simple, ballistic model. However, for a calibration of the non-local version of \threeeq\ and a better calibration of $\alpha_\omega$, 3D simulations or comparisons to observations are needed. We refer the reader to Sect.~\ref{sec:models} for a discussion of the values of the free parameters used to obtain the models discussed in this paper.

\section{Observations and Models}\label{sec:obsandmodels}

\subsection{Observations}\label{sec:obs}
The Sun is the ideal benchmark to test new ingredients in stellar models because of the unprecedented data quality we have due to its proximity. Helioseismology makes it possible to access the interior structure. Below, we give a brief overview of the observables used to compare with the solar models. They are the same as discussed by \citet{Braun2024}. For more details, we refer the reader to the respective section by \citet{Braun2024} and the review article by \citet{Christensen-Dalsgaard2021}, and references therein.

The solar sound speed profile is probably one of the most powerful observables for the Sun because it is derived from helioseismic inversions \citep{Basu2009}, which were found to be highly independent of the solar model used as reference model \citep{Basu2000}. 

The temperature gradient of a layer affects the sound speed within that layer, which can be measured by helioseismology. \citet{Basu1997b} used models with different depths of the convective envelope and minimized the difference between the model sound speed profile and that of the Sun. 
This way, the radius where the temperature gradient changes from close-to-adiabatic to radiative was found to be $R_\mathrm{cz}=0.713\pm 0.001$~\Rsun\ \citep{Basu1997b}. 

The variation in the adiabatic index caused by the second ionization zone of helium (He) can be used to measure the He abundance in the convective envelope \citep[$Y_\mathrm{cz} = 0.2485\pm0.0034$, ][]{Basu2004}.

The helioseismic surface effect is the difference between the observed frequencies and the frequencies calculated based on a stellar model. This difference is more pronounced for modes with higher frequencies because they probe layers closer to the surface. This systematic disagreement is caused by insufficient modelling of the convective surface layers in 1D stellar models (structural effect), as well as by the assumption that the oscillations are adiabatic (modal effect, \citealp{Houdek2017}). \citet{Jorgensen2019}, and \citet{Zhou2025} patched an averaged 3D atmosphere to a 1D stellar model to improve the modelling of these surface layers \citep[see also][]{Ball2016, Belkacem2019}. 
The patched models have a significantly reduced surface effect. In particular, the inclusion of turbulent pressure in the stellar model removed the structural effect almost completely.
In this paper, we will use their model without turbulent pressure for comparison, since our solar model  does not include turbulent pressure. This patched solar model, with an atmosphere from averaged 3D simulations but without considering turbulent pressure in the 1D interior, will be called the ``patched model'' hereafter.

\subsection{Solar Models}\label{sec:models}
We calculate solar calibrated models with different convection models and compare the interior structure to the observations described above. The stellar evolution code used in this work is GARSTEC\footnote{GARSTEC can be obtained on reasonable request from the authors, for more details, see \url{https://www.mpa-garching.mpg.de/84395/Structure-and-Evolution-of-Single-Stars}} which is described in detail by \citet{Weiss2008}.
We obtain solar calibrated models by adjusting the initial helium abundance, $Y_\mathrm{init}$, the initial metal abundance, $Z_\mathrm{init}$, and a free parameter of the convection theory. It is calibrated to reproduce the solar radius \Rsun, luminosity L$_\odot$ and surface metal-to-hydrogen ratio Z$_\odot/$X$_\odot$ at the age of the Sun. 
As described in \citet{Braun2024}, the solar model using MLT (\mlt) was obtained by adjusting the free parameter $\alpha_\mathrm{MLT}$ to match the solar properties to an accuracy of $\delta A/A_\odot\lesssim 2\cdot10^{-6}$, with $\delta A = A - A_\odot$ and $A \in \{R, L, Z/X\}$. The solar model using the 1-equation model (\ssmoneeq) was obtained by varying the equivalent free parameter $\alpha_\Lambda$ until an accuracy of $\delta A/A_\odot\lesssim 6 \cdot 10^{-5}$ was reached. All other free parameters of the convection theory were kept constant.

For the solar model with the 3-equation model (\ssmthreeeq), it is not straightforward to determine what value the free parameters should have. In Sect.~\ref{sec:kuhfuss}, we mentioned the standard choice for the free parameters, going back to estimates of \citet{Kuhfuss1987} using the local version of \threeeq. For the non-local version, \citet{Ahlborn2022} used $\alpha_\omega=\alpha_\Pi=\alpha_\Phi=0.3$ for convective cores of intermediate-mass main-sequence stars. They found a good agreement compared to models with classical MLT including overshooting, which were calibrated against observations. 
To obtain a calibrated solar model, we adjust $\alpha_\Pi$ and $\alpha_\Phi$, taking Kuhfuss' estimate for the local version of \threeeq\ as guidance, and because they are the parameters which affect the effective temperature and luminosity the most. After a first approximate adjustment, we kept $\alpha_\Phi=2.0$ constant and continued to vary $\alpha_\Pi$ to obtain a solar model with an accuracy $\delta A/A_\odot$ of $10^{-4}$ to $10^{-5}$. 
As outlined in Sect.~\ref{sec:kuhfuss}, varying the parameter $\alpha$ does not have the same effect as in MLT. In contrast, we found that changing it causes unphysical steps in the temperature gradient because of its influence on the dissipation by buoyancy waves (Eq.~\ref{eq:buoyancywaves}).

For the \ssmthreeeq\ described in Sect.~\ref{sec:solcalib}, all free parameters, except for $\alpha_\Pi$ and $\alpha_\Phi$, were kept at their default value. We acknowledge that this procedure introduces some ambiguity, and the effect of different values for the other free parameters will be investigated in Sect.~\ref{sec:varyparms}. 
We tested the effect of using $\alpha_\Pi=4.9$ and $\alpha_\Phi=3.3$ on convective cores using a 5~\Msun\ star. We found that even with these extreme values, the change is minor because a convective core is close-to-adiabatic, and thus, the temperature gradient is not greatly affected by the choice of parameters \citep[see also Appendix B from][]{Ahlborn2022}.
A thorough comparison with 3D hydrodynamical simulations is necessary to calibrate the free parameters of \threeeq\ \citep{Ahlborn2026}.

In all models, the OPAL equation of state \citep{Rogers2002} was used. In radiative regions, atomic diffusion was considered for hydrogen, helium, and metals.
The models were calibrated to reproduce Z$_\odot/$X$_\odot=0.0225$, following the composition of \citet{Magg2022}.
Here, our intention is not to study the composition itself, but the effects of a convection model different from MLT.  However, we include the solar model using the \threeeq\ and the \citet{Asplund2009}-abundances in Appendix~\ref{ap:asplund-abundances}. For a discussion of the effects of composition on the \oneeq, see \citet{Braun2024}.
The chemical composition used to calculate the opacities is always consistent with the respective solar composition. We used the OP opacities \citep{Badnell2005}, substituted with low temperature opacities \citep{Ferguson2005} (Yago Herrera, private communication).

\section{Results}\label{sec:results}

This section is divided into three parts. First, we present the results of the fiducial solar calibrated model using the \threeeq\ in Sect.~\ref{sec:solcalib}. This model was obtained by adjusting $\alpha_\Pi$, while keeping $\alpha_\Phi=2.0$ and the other free parameters of \threeeq\ at their default values. In Sect.~\ref{sec:varyparms}, we study the influence of the free parameters of \threeeq. Due to findings from Sect.~\ref{sec:solcalib}, we test the effect of the local closure relations (Eq.~\ref{eq:local-closure}) on the outermost layers in Sect.~\ref{sec:frankenstein}.

\subsection{Solar Model with the 3-equation Model}\label{sec:solcalib}

The free parameters used to obtain the model discussed in this section (the ``fiducial model'') are given in Table~\ref{tab:vary-parms}. In addition, Table~\ref{tab:vary-parms} states the achieved accuracy of the luminosity, radius and metal-to-hydrogen ratio, and the helium abundance and depth of the convective zone ($Y_\mathrm{cz}$ and $R_\mathrm{cz}$). The \mlt\ and \ssmoneeq\ were already discussed in detail by \citet{Braun2024}.

\subsubsection{The Inner Boundary}\label{sec:lower-boundary}

\begin{figure}
    \centering
    \includegraphics[width=1.0\linewidth]{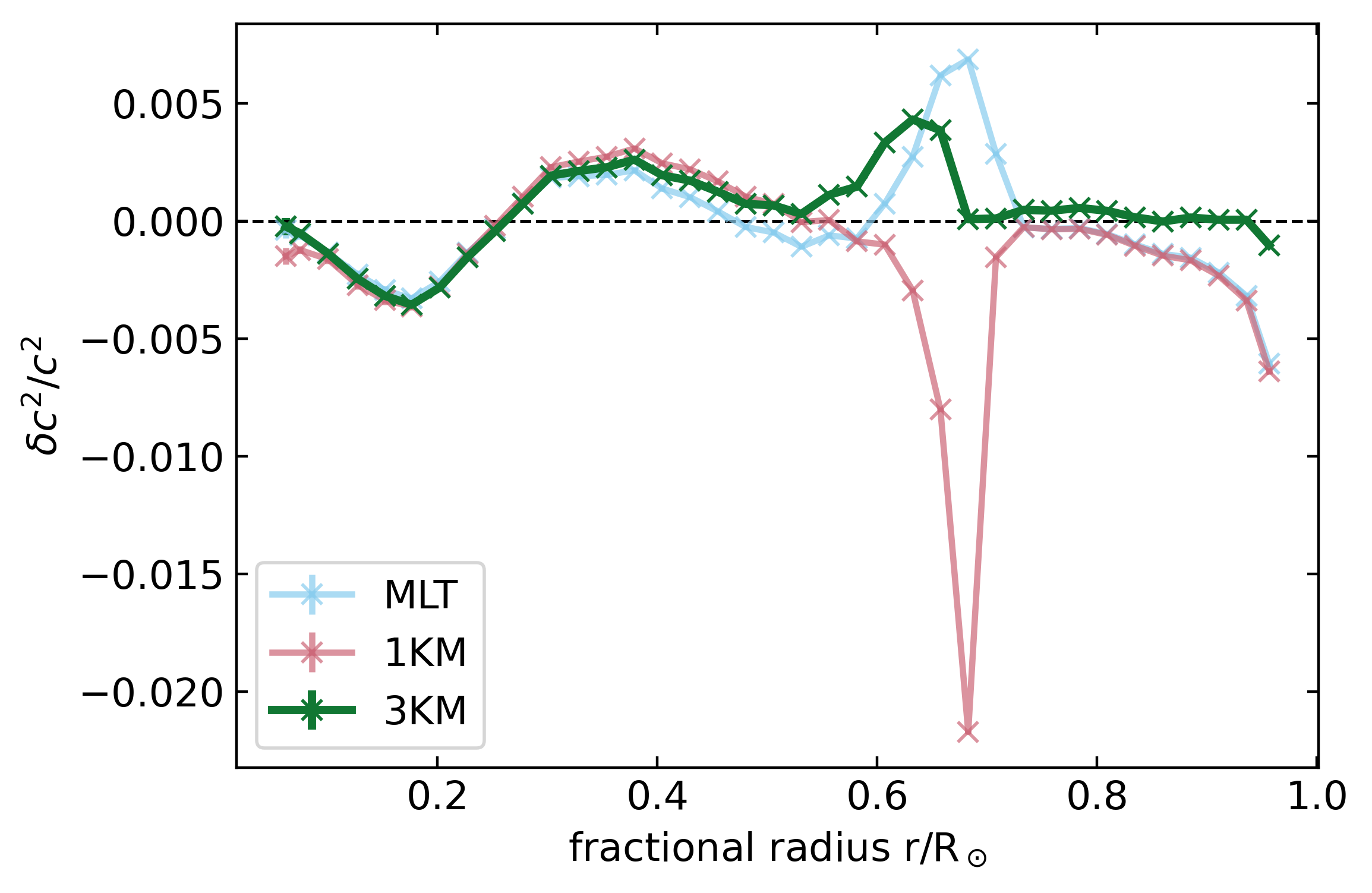}
    \caption{The relative difference of the squared sound speed of the helioseismic measurement $c_\mathrm{helio}$ and the solar calibrated models $c_\mathrm{model}$, using MLT (light blue), \oneeq\ (light pink), and \threeeq\ (green). The quantity on the y-axis is defined as: $\delta c^2/c^2 = (c^2_\mathrm{helio}-c^2_\mathrm{model})/c^2_\mathrm{helio}$. The crosses denote the radii of the data points from the helioseismic inversion.}
    \label{fig:soundspeed}
\end{figure}

The treatment of convection has a direct effect on the sound speed, which can be measured by helioseismology and can be used to assess the interior structure. Figure~\ref{fig:soundspeed} shows the relative squared sound speed difference between the helioseismic measurement $c_\mathrm{helio}$ \citep{Basu2009} and the model $c_\mathrm{model}$: $\delta c^2/c^2 = (c^2_\mathrm{helio}-c^2_\mathrm{model})/c^2_\mathrm{helio}$. The result in the radiative interior is similar to what is obtained when using MLT or \oneeq, as expected. 
The most notable difference between the models is the reduction of $\delta c^2/c^2$ below the boundary of the convective region at approximately $r=0.65$~\Rsun. 
In this region, \threeeq\ reduces the difference by 37\% compared to \mlt\ and 80\% compared to \ssmoneeq.
Since the sound speed profile is directly affected by the temperature stratification, this is linked to the predicted temperature gradient in the CBM region.
At radii $\gtrapprox 0.8$~\Rsun, the sound speed profile is strongly affected by the accuracy of the calibrated radius, where a small improvement can have a significant effect. Therefore, the effects of the convection theory and the calibration are difficult to disentangle in this region (see also Appendix~\ref{ap:test-calibration}).

\begin{table*}[htb]
\caption{Solar models with different parameter combinations (Sect.~\ref{sec:varyparms}), and different treatments of the outer layers (Sect.~\ref{sec:frankenstein}).}
\label{tab:vary-parms}
    \centering
    \begin{tabular}{c|c|ccc|cc}
    \hline\hline
       Name & $\alpha_\Pi$  & $\delta R/$R$_\odot$ & $\delta L/$L$_\odot$ & $\delta (Z/X)$/(Z$_\odot$/X$_\odot$) & $Y_\mathrm{cz}$  & $R_\mathrm{cz}$  \\
         &    & [$10^{-4}$] & [$10^{-4}$] & [$10^{-4}$] &    & [\Rsun]  \\       
    \hline
    fiducial               & 2.155  &  0.66  & -0.89 &  3.3  & 0.2455 & 0.7067 \\ \hline
    $\alpha_\Phi=1.0$      & 2.8    & -11  &  1.6  & -11   & 0.2453 & 0.7059 \\
    $\alpha_\Phi=3.0$      & 1.9    &  12  & -5.3  &  12   & 0.2457 & 0.7077 \\
    $\alpha_\omega=0.1 $   & 2.1    &  5.7   & -2.1  & -75   & 0.2442 & 0.7111 \\
    $\alpha_\omega=0.5 $   & 2.21   & -3.3   & -5.2  &  68   & 0.2467 & 0.7025 \\
    $C_{D}=1.0$            & 2.39   &  11  & -6.1  &  34   & 0.2461 & 0.7054 \\
    $C_{D}=3.4$            & 2.12   & -3.6   &  0.25 & -15   & 0.2452 & 0.7072 \\ 
    $\gamma=1.7$           & 4.7    & -2.1   &  1.9  &  12   & 0.2457 & 0.7062 \\ 
    $\gamma=5.2$           & 0.9    &  0.38  & -0.11 & -2.2  & 0.2454 & 0.7067 \\ 
    $c_4=0.02$             & 2.155  &  -3.0  & -0.52 &  153  & 0.2482 & 0.7011 \\ 
    $c_4=0.2$              & 2.155  & 2.8    & -1.2  & -100  & 0.2438 & 0.7103 \\ \hline
    Case A                 & 2.88   & -2.0   &   2.1 & -33   & 0.2452 & 0.7066  \\
    Case B                 & 3.47   & -1.9   &  -2.4 &  2.1  & 0.2455 & 0.7061  \\
    Case C                 & 6.76   & -2.7   &  -5.6 & -2.6  & 0.2454 & 0.7056  \\
    \hline
    \multicolumn{7}{l}{\small The parameter values for the fiducial model are $\alpha_\Phi=2.0$, $\alpha_\omega=0.25$, $C_D=2.177$, $\gamma=3.46$, $c_4=0.072$.} \\
    \multicolumn{7}{l}{\small  The first column specifies the changed parameter and its value if different from the fiducial model.} \\
    \end{tabular}
\end{table*}

\begin{figure}
    \centering
    \includegraphics[width=1.0\linewidth]{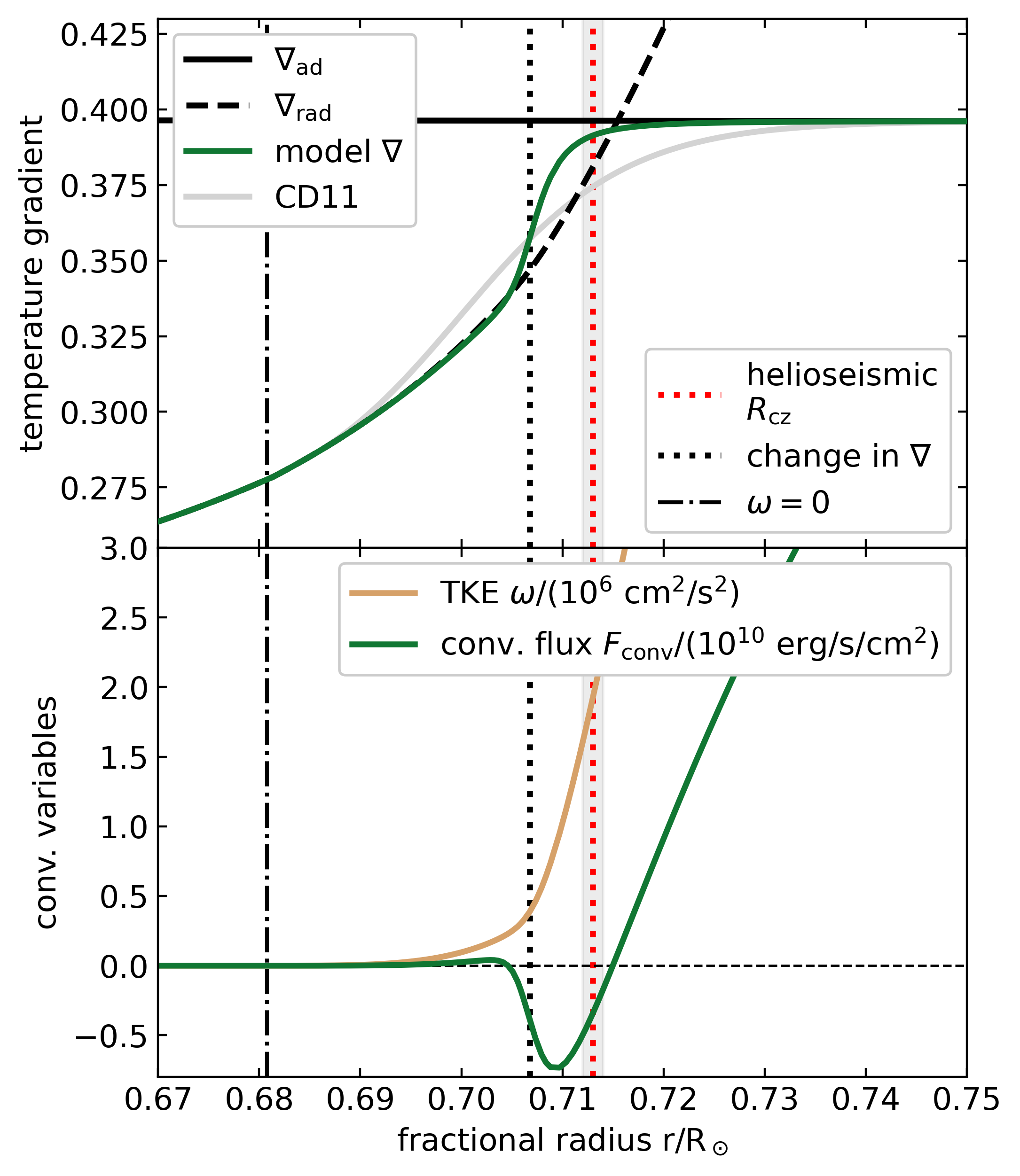}
    \caption{The temperature gradient of the \ssmthreeeq\ (green) at the inner boundary of the convective envelope in comparison with the adiabatic (black, solid line), and radiative (black, dashed line) temperature gradient. 
    The boundary of the convective region as predicted by the model (black, dotted line) and as measured by helioseismology (red, dotted line) are indicated by vertical lines. The gray shaded regions indicate the 1$\sigma$ uncertainty of the measurement. The dash-dotted line denotes the radius where $\omega=0$.
    The light gray line indicates the temperature profile deduced by \citet{Christensen-Dalsgaard2011}.}
    \label{fig:temp.grad}
\end{figure}

The upper panel of Fig.~\ref{fig:temp.grad} shows the temperature gradient at the inner boundary of the solar envelope of \ssmthreeeq. The temperature gradient of the model ($\nabla$) is close to the adiabatic temperature gradient ($\nabla_\mathrm{ad}$) in the layers where $\nabla_\mathrm{ad}$ is less than the radiative temperature gradient ($\nabla_\mathrm{rad}$). However, different from \mlt\ and \ssmoneeq, the temperature gradient of \ssmthreeeq\ is weakly subadiabatic at radii of 0.716 to 0.840~\Rsun\ ($-2\cdot 10^{-3}<\nabla-\nabla_\mathrm{ad}<-6.5\cdot 10^{-8}$, median: $-2\cdot 10^{-5}$), instead of being weakly superadiabatic. Although $\nabla<\nabla_\mathrm{ad}$, the convective flux $F_\mathrm{conv}$ is positive (Fig.~\ref{fig:temp.grad}, lower panel). This is what defines a Deardorff layer \citep{Deardorff1966} and is known from 3D simulations \citep{Kapyla2017, Kapyla2025} and atmospheric sciences \citep[][and references therein]{Deardorff1966}.
Starting near the Schwarzschild boundary, $\nabla$ smoothly changes from close to $\nabla_\mathrm{ad}$ to $\nabla_\mathrm{rad}$. In the layer where $\nabla_\mathrm{rad}<\nabla<\nabla_\mathrm{ad}$, $F_\mathrm{conv}$ is negative. Below this layer, $F_\mathrm{conv}$ becomes positive again while $\nabla\approx\nabla_\mathrm{rad}$ and the TKE $\omega>0$. The value of $F_\mathrm{conv}$ in this region is two orders of magnitudes smaller compared to the bulk of the convective zone (max. $4.0\cdot 10^{8}$~erg/s/cm$^2$ compared to max. $8.0\cdot 10^{10}$~erg/s/cm$^2$, median $6.4\cdot 10^{10}$~erg/s/cm$^2$).

\citet{Christensen-Dalsgaard2011} investigated what kind of temperature gradient fits the helioseismic data best in the CBM region by parametrizing and varying the slope of $\nabla$, without an underlying convection theory. They found that a smooth transition from $\nabla_\mathrm{ad}$ to $\nabla_\mathrm{rad}$ in the CBM region and a subadiabatic stratification in the lower convective zone fits the data best. This optimised temperature profile is included in Fig.~\ref{fig:temp.grad} (grey, solid line). While there are differences in the detailed shape of the temperature gradients derived by \citet{Christensen-Dalsgaard2011} and obtained with \threeeq, the overall features of a more smooth transition and a subadiabatic stratification already within the Schwarzschild unstable zone agree. This supports the temperature stratification of the \ssmthreeeq\ qualitatively.

\begin{figure}
    \centering
    \includegraphics[width=1.0\linewidth]{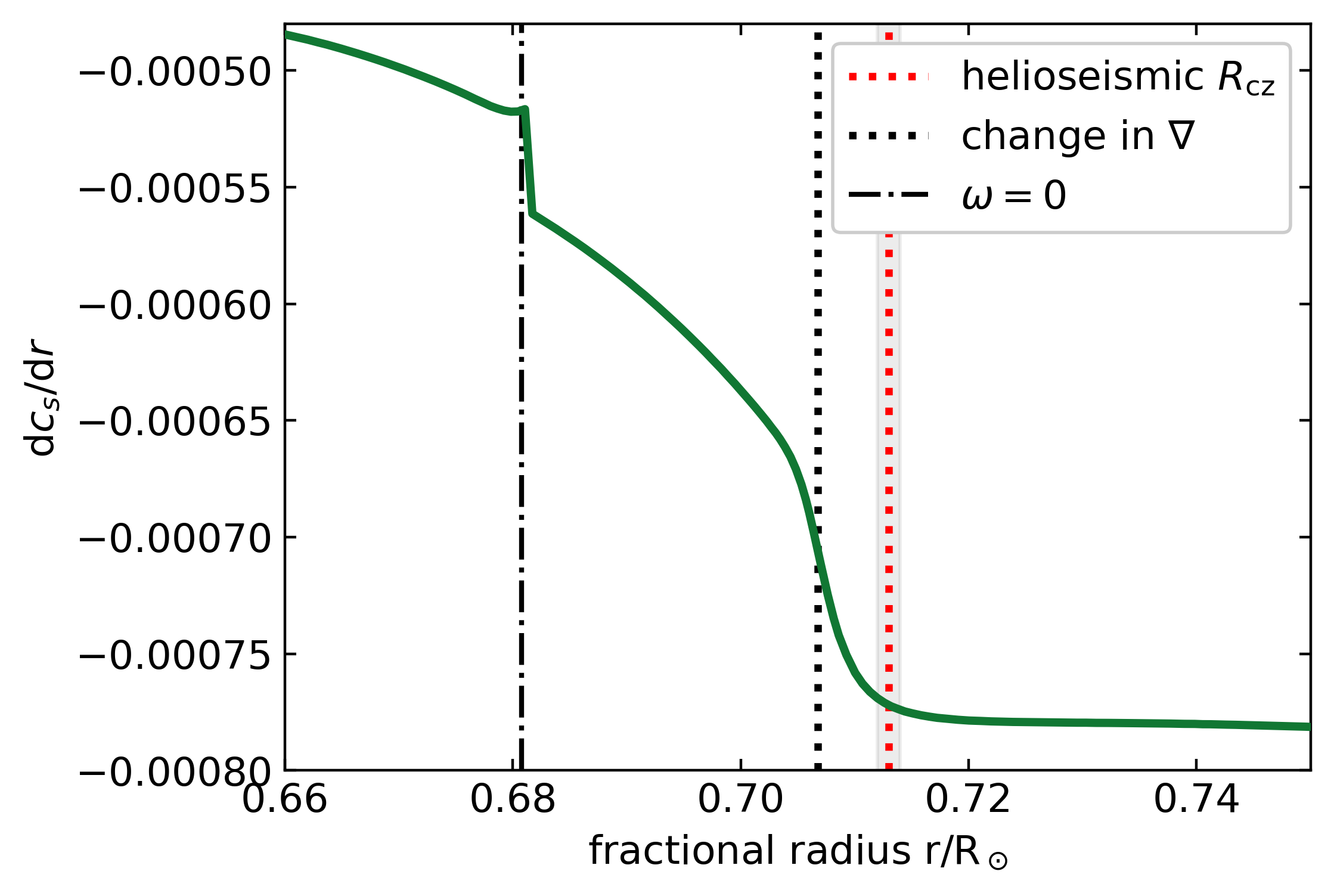}
    \caption{The derivative of the sound speed of the \ssmthreeeq. The vertical lines denote the boundary of the convective envelope as measured by helioseismology (red, dotted), and as predicted by the \ssmthreeeq\ (black, dotted). The dash-dotted vertical line denotes the radius where $\omega=0$. The 1$\sigma$ ranges of the measurement are indicated by the gray shaded region.}
    \label{fig:csderivs}
\end{figure}

The vertical lines in Fig.~\ref{fig:temp.grad} denote the radius where the TKE becomes zero (dash-dotted line), and the radius where the change in the slope of $\nabla$ is largest (black, dotted line). Both these features leave their imprint in the sound speed profile (see Fig.~\ref{fig:csderivs}) and are expected to influence the frequencies of the solar oscillations.
The transition from a close-to-adiabatic to a radiative temperature gradient (Fig.~\ref{fig:temp.grad}), and its effect on the sound speed profile (Fig~\ref{fig:csderivs}) occurs at a radius of 0.7067~\Rsun. This lies within 6.3$\sigma$ of the helioseismic measurement of the depth of the convective envelope \citep[$0.713\pm0.001$, ][]{Basu1997b}. This is a clear improvement compared to \ssmoneeq\ \citep[29$\sigma$, ][]{Braun2024}. 
We note, that \citet{Basu1997b} obtained this measurement using convection models that had either no overshooting or overshooting assuming a close-to-adiabatic temperature gradient in the CBM region, which may influence the inferred value for $R_\mathrm{cz}$. A new determination based on \ssmthreeeq\ would be necessary.
The second feature, which can be seen in Fig.~\ref{fig:csderivs}, is at a radius of 0.6807~\Rsun, which is the same radius where the TKE becomes zero, and the mixing of elements stops. This feature is caused by the dependence of the sound speed on molecular weight. However, helioseismology does not detect such a second glitch \citep{Basu1997b}. A more gradual change in molecular weight, resulting from partial mixing, may help to remove this feature.

The He abundance of the envelope of \ssmthreeeq\ ($Y_\mathrm{cz}=0.2455$, see Table~\ref{tab:vary-parms}) is in 0.9$\sigma$ agreement to the one measured by helioseismology \citep[$Y_\mathrm{cz}=0.2485\pm0.0034$, ][]{Basu2004}. In short, the \threeeq\ with standard parameters agrees much better with the sound speed profile of the Sun than the standard MLT model, and it agrees very well in $Y_\mathrm{cz}$ with the seismic value. \threeeq\ improves $R_\mathrm{cz}$ compared to \oneeq, and deviates by 6.3$\sigma$ from the value determined by \citet{Basu1997b}, which itself may need to be redetermined by using solar models with a non-local convection theory. Based on the sound speed profile of the \threeeq, two glitches, caused by the change in molecular weight and the temperature gradient, would be expected which does not agree with helioseismology.

\subsubsection{The Outer Boundary}\label{sec:upper-boundary}

\begin{figure}
    \centering
    \includegraphics[width=1.0\linewidth]{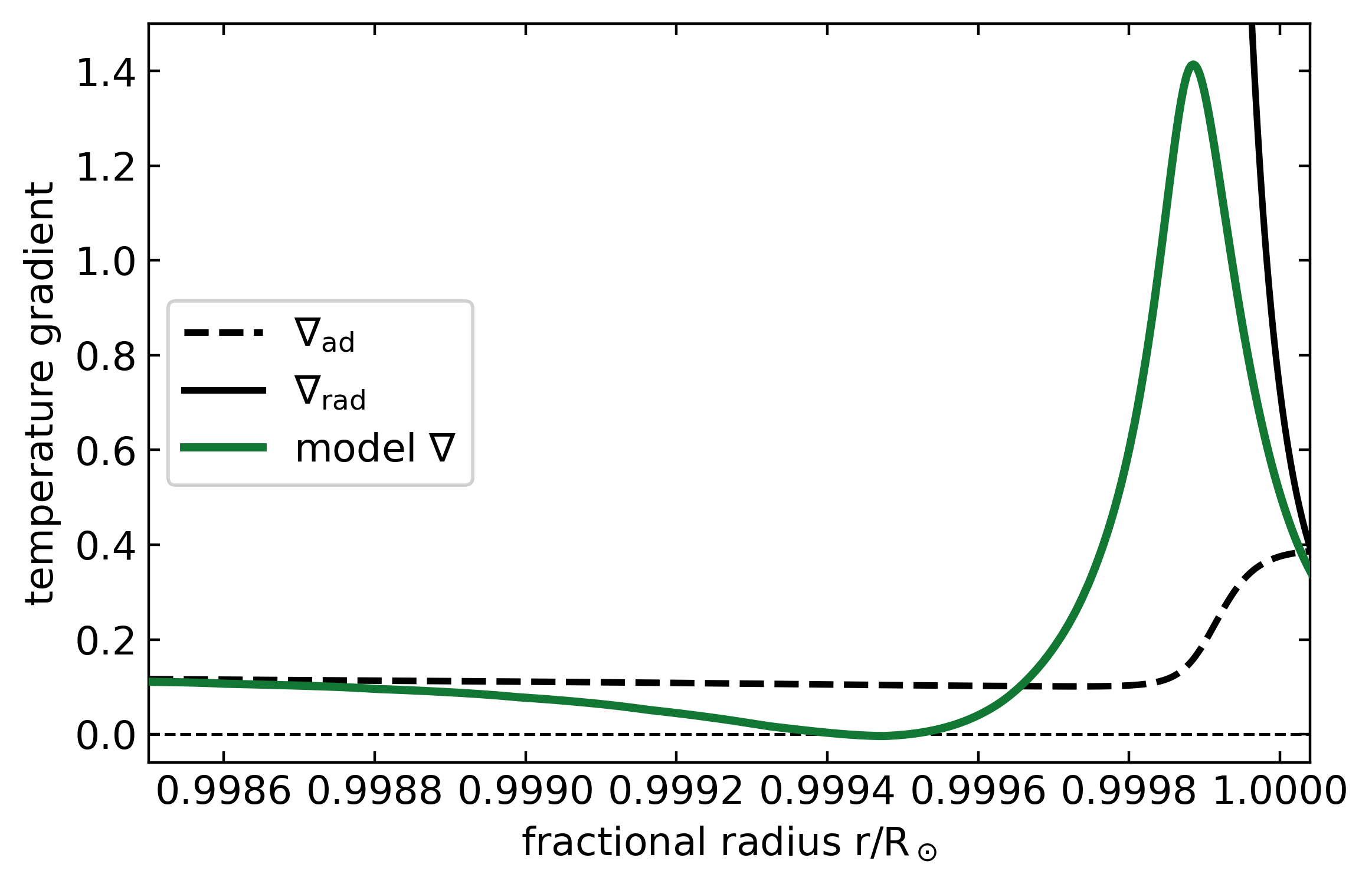}
    \caption{Temperature stratification of the outer 0.0015~\Rsun\ of the \ssmthreeeq.}
    \label{fig:temp-grad-outerbound}
\end{figure}

\begin{figure}
    \centering
    \includegraphics[width=1.0\linewidth]{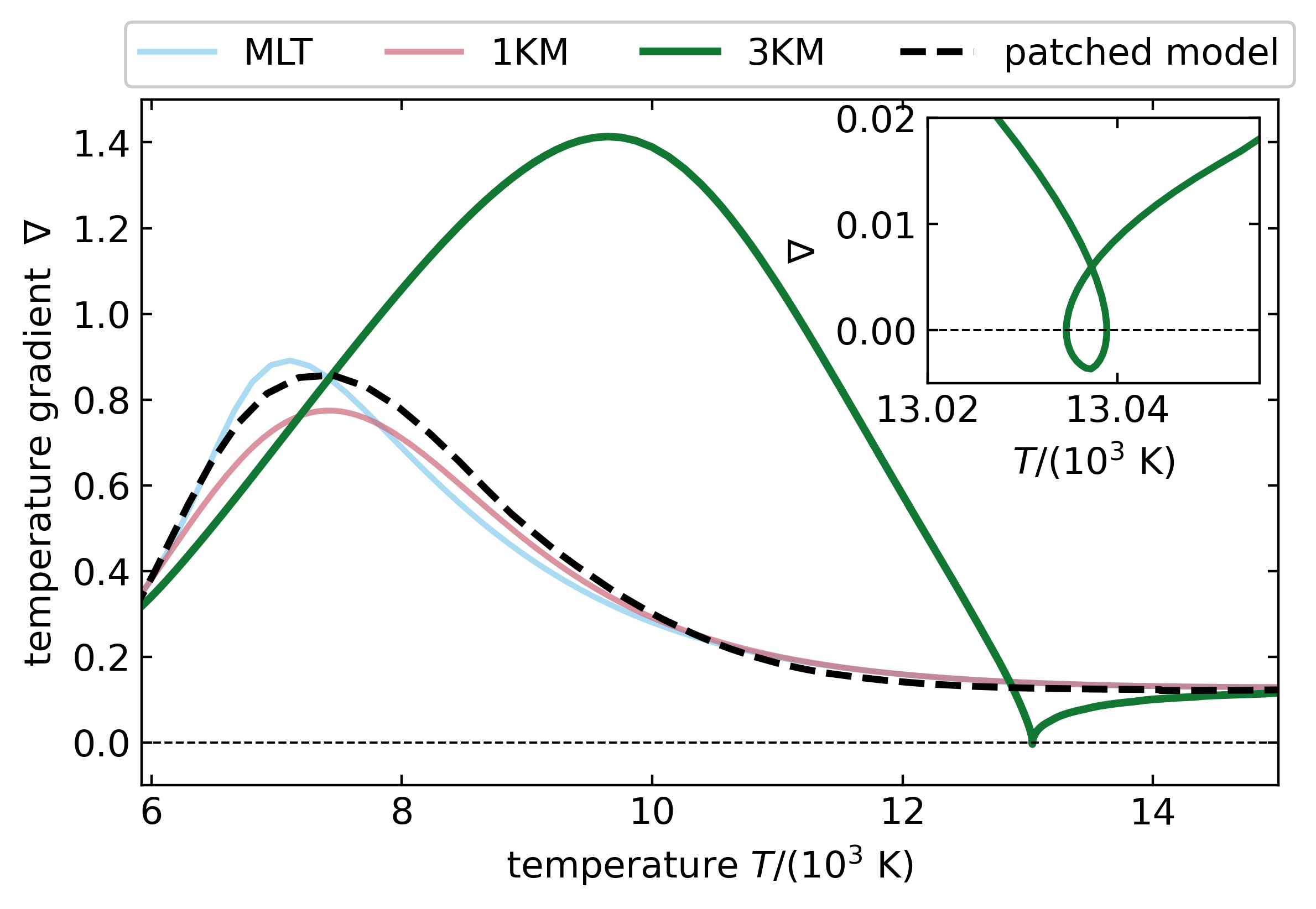}
    \caption{The temperature gradient against the temperature of \mlt\ (light blue), \ssmoneeq\ (light pink), and \ssmthreeeq\ (green). The thick dashed line shows the temperature gradient of the patched model \citep{Jorgensen2019}. The temperature range which is shown corresponds to the outermost 0.0017~\Rsun. The inset shows a zoom-in into the region with a negative temperature gradient.}
    \label{fig:gradT-T}
\end{figure}

The profiles of the temperature gradient at the upper boundary of the convective zone, that means right below the surface, predicted by MLT and \oneeq, are very similar.
They predict a weakly superadiabatic temperature gradient in the bulk of the convective region, which becomes strongly superadiabatic close to the surface (MLT: $\nabla-\nabla_\mathrm{ad} = 0.55$ at $r/\mathrm{R}_\odot=0.99992$; \oneeq: $\nabla-\nabla_\mathrm{ad} = 0.46$, at $r/\mathrm{R}_\odot=0.99991$). 
In the \ssmthreeeq\ this superadiabatic layer (SAL) has a stronger superadiabaticity of $\nabla-\nabla_\mathrm{ad} = 1.25$ (at $r/\mathrm{R}_\odot$=0.99989, see Fig.~\ref{fig:temp-grad-outerbound}). Towards smaller radii, the temperature gradient becomes subadiabatic, even negative (0.9981~\Rsun\ to 0.9997~\Rsun, minimum $\nabla=-3.7\cdot10^{-3}$), before it becomes weakly superadiabatic (0.8394~\Rsun\ to 0.9981~\Rsun, median: $\nabla-\nabla_\mathrm{ad} = 2.2\cdot 10^{-4}$).
We use hydrodynamical simulations of the solar atmosphere from the literature to compare the temperature stratification in the outer layers. Figure~\ref{fig:gradT-T} shows the temperature gradient against temperature. The temperature inversion in \ssmthreeeq\ is shown in detail in the inset in this figure, which zooms in on the relevant region. This can be compared to 
the patched model, which uses the averaged 3D simulation from the Stagger grid \citep[][black, dashed line in Fig.~\ref{fig:gradT-T}]{Magic2013, Jorgensen2019, Zhou2025}. A layer with a negative temperature gradient is clearly not present in these 3D simulations, and neither in the 2D simulations presented in \citet[][their Fig.~2]{Schlattl1997}.

\begin{figure}
    \centering
    \includegraphics[width=1.0\linewidth]{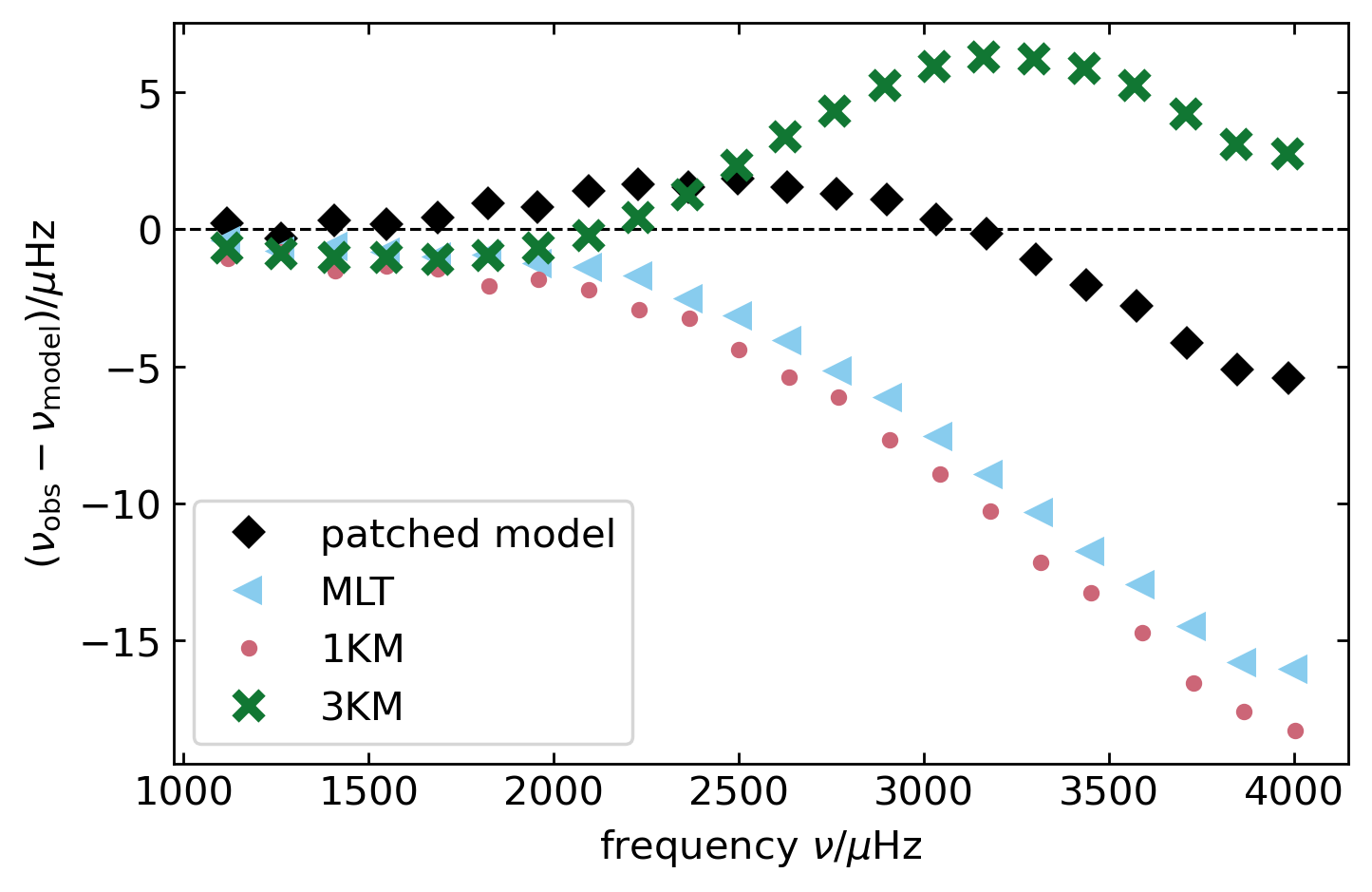}
    \caption{The helioseismic surface effect, that is the mismatch between the observed frequencies and the ones calculated from \mlt\ (light blue), \ssmoneeq\ (light pink), \ssmthreeeq\ (green), and the solar model using an averaged 3D atmosphere as outer boundary condition \citep[``patched model'', black, ][]{Jorgensen2019}.}
    \label{fig:surface-effect}
\end{figure}

The outer layers also affect the helioseismic surface effect. Using GYRE \citep{Townsend2013}, we calculated the frequencies of the solar models with the different convection theories. We found an improvement in the surface effect (Fig.~\ref{fig:surface-effect}). The maximum absolute difference between the observed and modelled frequencies decreases from 16.0~$\mu$Hz to 6.3~$\mu$Hz, when applying the \threeeq, in comparison to MLT (\oneeq: 18.3~$\mu$Hz).
\citet{Demarque1997, Demarque1999} found that a stronger superadiabaticity in the SAL reduces the surface effect, which is in agreement with the properties of the \ssmthreeeq\ (Fig.~\ref{fig:gradT-T}). 
However, the patched model \citep{Jorgensen2019, Zhou2025} improves the surface effect without having a stronger superadiabaticity in the SAL compared to \mlt\ and \ssmoneeq.
This suggests that the surface effect is not only dependent on the superadiabaticity of the SAL.

\begin{figure}
    \centering
    \includegraphics[width=1.0\linewidth]{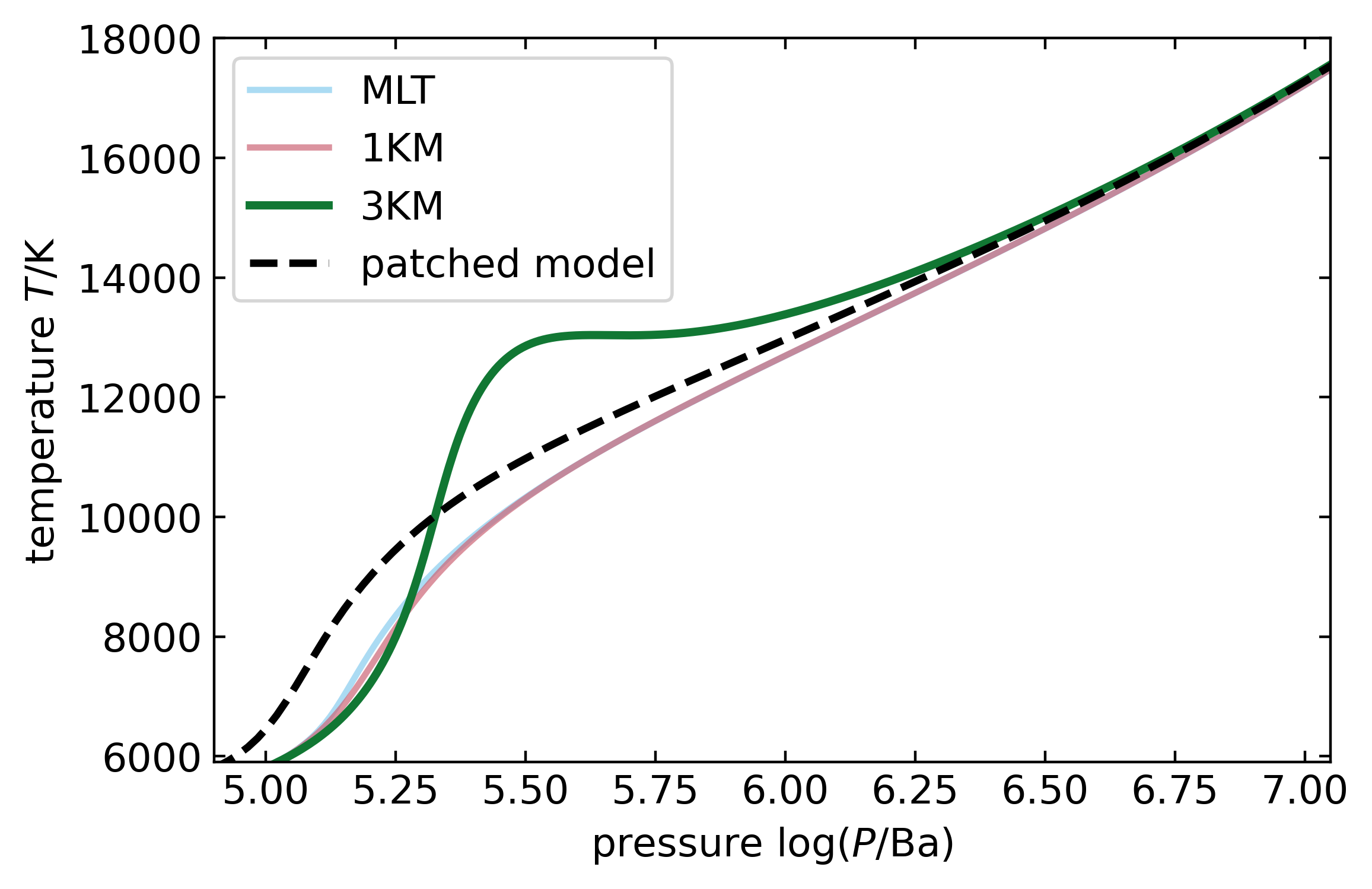}
    \caption{Pressure against the temperature in the uppermost regions. The light pink line shows the profile for \ssmoneeq, which largely overlaps with the light blue line showing the \mlt. The black, dashed line shows the profile of the solar model using an averaged 3D atmosphere as outer boundary condition \citep[``patched model'', ][]{Jorgensen2019}. \ssmthreeeq\ is represented by the green line. The segment which is shown corresponds to the outermost 0.0026~\Rsun.}
    \label{fig:T-P-outerbound}
\end{figure}

Figure~\ref{fig:T-P-outerbound} shows pressure against temperature in the uppermost layers. The patched model has a higher temperature for a given pressure for $\log{P/\mathrm{Ba}}\lessapprox6.5$ compared to \ssmoneeq\ and \mlt. The higher temperature at a given pressure in this range seems to be a common characteristic of the patched model \citep{Jorgensen2019, Zhou2025}, the 2D simulation \citep{Schlattl1997}, and the \ssmthreeeq, which all improve the surface effect but have otherwise different relations of pressure and temperature and different profiles of $\nabla$ in the SAL.

Finally, the structure of the SAL does also affect the effective temperatures of the models. The solar models have the same effective temperature by construction. But the effect on the effective temperatures becomes relevant in the later evolutionary stages.
Starting from the solar calibrated models, we continue the evolution up the RGB. Using the \oneeq\ instead of MLT causes only minor differences in the effective temperatures along the RGB evolution.
The model using \threeeq, in contrast, evolves at lower effective temperatures than the ones using \oneeq\ or MLT. 
A more detailed study is needed to clarify if the shift in effective temperature as seen when using \threeeq\ is favourable or not (see also Fig.~\ref{fig:hrd}, and Sect.~\ref{sec:disc:literatur}).

In summary, the \threeeq\ includes non-local effects without any ad hoc description, different from convection models assuming instant mixing over a fraction of the pressure scale height. It improves the temperature stratification at the inner boundary of the convective region. However, especially at the outer boundary, features arise which are unrealistic (particularly the local temperature inversion around $T \sim 13\cdot 10^3~\mathrm{K}$).
Next, we therefore study whether choosing different model parameters can remove these discrepancies and improve the solar model. Furthermore, in Sect.~\ref{sec:frankenstein}, we test the effect of the local closure relations of the $\Pi$- and $\Phi$-equation.

\subsection{Effects of Varying Free Parameters}\label{sec:varyparms}

The \ssmthreeeq\ discussed in Sect.~\ref{sec:solcalib} (``fiducial model'') used a specific set of the free parameters of the \threeeq. However, other parameter combinations can also result in a solar calibrated model. To test the effect of the different parameters on the interior structure, we changed the value of one parameter at a time and calibrated the model by varying $\alpha_\Pi$, as before. Table~\ref{tab:vary-parms} gives an overview of the parameters used, the resulting values for $\alpha_\Pi$, $\delta R$/R$_\odot$, $\delta L$/L$_\odot$, and $\delta (Z/X)$/(Z$_\odot$/X$_\odot$), as well as for the He content $Y_\mathrm{cz}$ and the depth $R_\mathrm{cz}$ of the convective envelope. 

All sets of parameters result in solar models which fit the helioseismic measurement of the helium abundance in the convective envelope by $<1.4\sigma$ \citep[Y$_\mathrm{cz}=0.2485\pm0.0034$,][]{Basu2004}. For models with a deeper convective envelope, the agreement is better.

\begin{figure*}[htb!]
    \centering
    \includegraphics[width=1.0\linewidth]{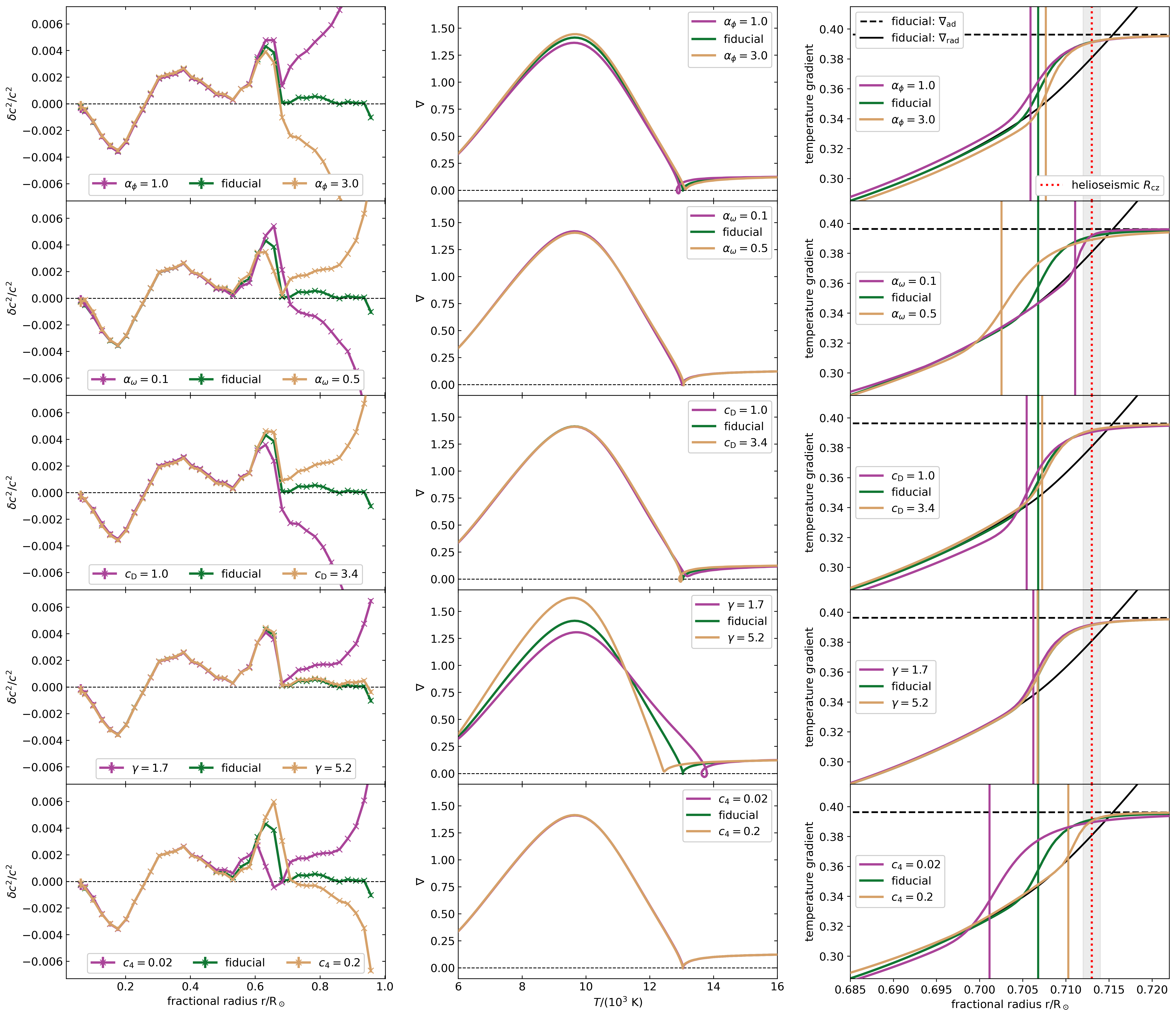}
    \caption{The solar models obtained with the \threeeq\ and different sets of free parameters (see Table~\ref{tab:vary-parms}). From left to right, the panels show the difference of the sound speed profile between models and observations, the temperature stratification of the outer layers, and the temperature stratification at the inner boundary of the convective envelope. The vertical lines in the right panels denote the radius where the change of the slope in the profile of the temperature gradient is strongest. The red, dotted line denotes the helioseimisc measurement, with the 1$\sigma$ ranges indicated by the gray shaded regions.}
    \label{fig:varyparams}
\end{figure*}

The left panels in Fig.~\ref{fig:varyparams} show the sound speed profiles of the models with different combinations of free parameters. Even when ignoring the outer layers, which are strongly affected by the exact radius of the model (see Appendix~\ref{ap:test-calibration}), differences can be seen at the inner boundary of the convective region.
Investigating the profile of the temperature gradient shows that the slope of the temperature gradient in the CBM region is correlated with the prominent feature around 0.65~\Rsun: The steeper the change in $\nabla$ (see right panels of Fig.~\ref{fig:varyparams}), the larger the difference between model and observations in the sound speed profile in this region. This supports the finding of \citet{Christensen-Dalsgaard2011} that a smooth change from $\nabla \approx \nabla_\mathrm{ad}$ to $\nabla_\mathrm{rad}$ gives the best agreement with helioseismic data.

The base of the convective envelope, marked with vertical lines in the right panels of Fig.~\ref{fig:varyparams}, shows the opposite. If the change of $\nabla$ to $\nabla_\mathrm{rad}$ is more gradual, $R_\mathrm{cz}$ is at smaller radii, and therefore in stronger disagreement with the helioseismic measurement by \citet{Basu1997b} of $0.713\pm0.001$~R$_\odot$. 

The models which show a larger improvement in the sound speed profile do not necessarily give a better result at the surface layers (middle panels in Fig.~\ref{fig:varyparams}). For example, the model with $\alpha_\Phi=1.0$ results in a very smooth change from $\nabla \approx \nabla_\mathrm{ad}$ to $\nabla_\mathrm{rad}$, but the temperature gradient becomes more negative at $T\approx 13 \cdot 10^{3}$~K compared to the fiducial parameter combination, or most of the other models. Furthermore, in all cases, the subadiabatic region between the SAL and the second superadiabatic region persists.

The parameters can be classified based on the region they influence most. The parameters $\alpha_\omega$ and $c_4$ mainly influence the inner boundary of the convective region. This is because $\alpha_\omega$ controls the non-local part of the TKE equation and $c_4$ controls the dissipation by buoyancy waves, which only affects the CBM region. The parameter $\gamma$ has only a minor effect on the inner boundary of the convective region, its main effect is at the outer boundary, where radiative losses become significant. The other parameters, $\alpha_\Phi$ and $C_\mathrm{D}$, influence both boundaries of the convective region.

In summary, we conclude that the unrealistic, strongly subadiabatic temperature gradient below the superadiabatic peak is likely not an artifact of a bad parameter choice but a sign that the convection model itself needs further improvement. Apart from this, comparisons to 3D simulations are needed to study and calibrate the free parameters of the \threeeq, since the effect of the individual parameters on the structure is too intertwined to calibrate them based on observational data.

\subsection{Alternative Closure Relations for $\Pi$ and $\Phi$}\label{sec:frankenstein}

To test where the unrealistic layer below the SAL comes from, we used the local closure relations for the $\Pi$- and $\Phi$-equations (Eq.~\ref{eq:local-closure}) for $r/\mathrm{R}_\odot \gtrapprox 0.979$, instead of the non-local ones, which we continue to use for $r/\mathrm{R}_\odot \lessapprox 0.979$ (Eq.~\ref{eq:non-local-closure}). We distinguish three cases: Case A: the TOMs of the equations for $\Phi$ and for $\Pi$ are both modelled with the local closure relation; Case B (Case C): we use the local closure relation for $\Phi$ ($\Pi$) and the non-local closure relation for $\Pi$ ($\Phi$). See Table~\ref{tab:cases} for an overview of the different cases. For all cases, $\alpha_\Pi$ needed to be adjusted to obtain a solar model with the correct radius and luminosity. All other free parameters have the same values as in the fiducial model. Table~\ref{tab:vary-parms} shows the values for $\alpha_\Pi$ and the achieved accuracy.

\begin{table}[htb!]
    \caption{Treatment of the TOMs of the $\Pi$- and $\Phi$-equation of the different cases}
    \label{tab:cases}
    \centering
    \begin{tabular}{c|cc}
    \hline\hline
                & $\Pi$     & $\Phi$ \\ \hline
         Case A & local     & local \\
         Case B & non-local & local \\
         Case C & local     & non-local \\ \hline
    \end{tabular}
\end{table}

\begin{figure}
    \centering
    \includegraphics[width=1.0\linewidth]{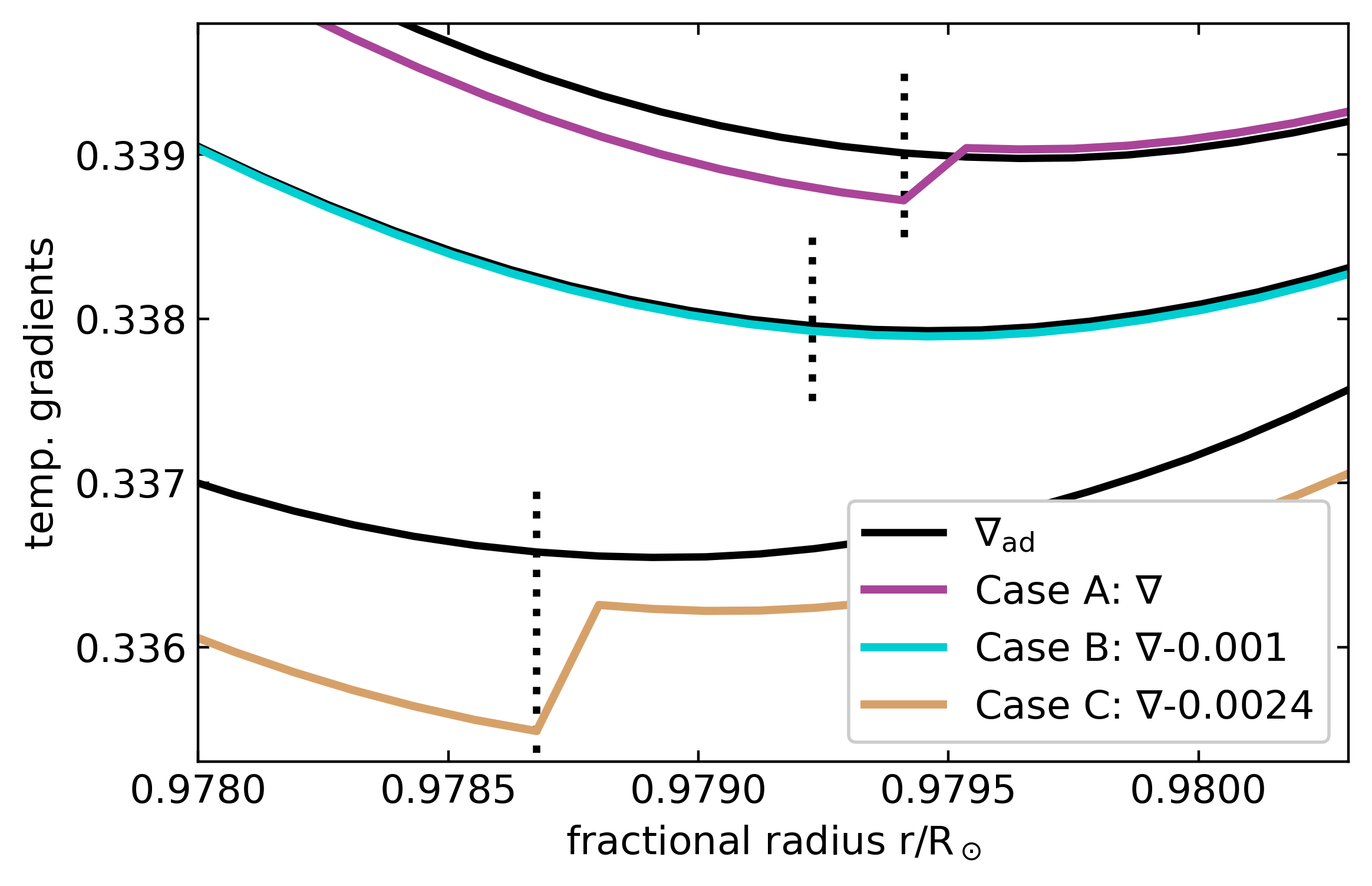}
    \caption{The radius at which the models switch to the local closure relations for the outer regions. Case A is shown in purple. For better visibility, case B (turquoise) and case C (orange) are shifted by -0.001 and -0.0024, respectively (see Table~\ref{tab:cases}). The black lines show $\nabla_\mathrm{ad}$ of the different models, shifted by the same amount as the model's $\nabla$. The dotted, vertical lines mark the last grid point at which the fully non-local \threeeq\ was applied.}
    \label{fig:switchpoint}
\end{figure}

Switching to local closure relations in the outer layers of the model clearly introduces inconsistency into the solar model. At the switch point, the shift from the fully non-local \threeeq\ to a version with at least one local closure relation causes a jump in the temperature gradient of varying degree (see Fig.~\ref{fig:switchpoint}), the largest one in case C. For case A, $\nabla$ jumps from being sub- to superadiabatic at the switch point. For case B the jump is smallest. However, these calculations are only meant as a test to understand better which term is causing the unrealistic behaviour of the outermost layers. The effect on the inner boundary of the convective envelope when switching to the local closure relations in the outermost layers is only minor.
For all cases, the temperature gradient does not become negative any more. Since the different treatment of the closure relations is only applied to the outermost layers, the subadiabatic region between the SAL and a second superadiabatic region persists. The details of the profiles of the temperature gradients are different between the cases.

\begin{figure}
    \centering
    \includegraphics[width=1.0\linewidth]{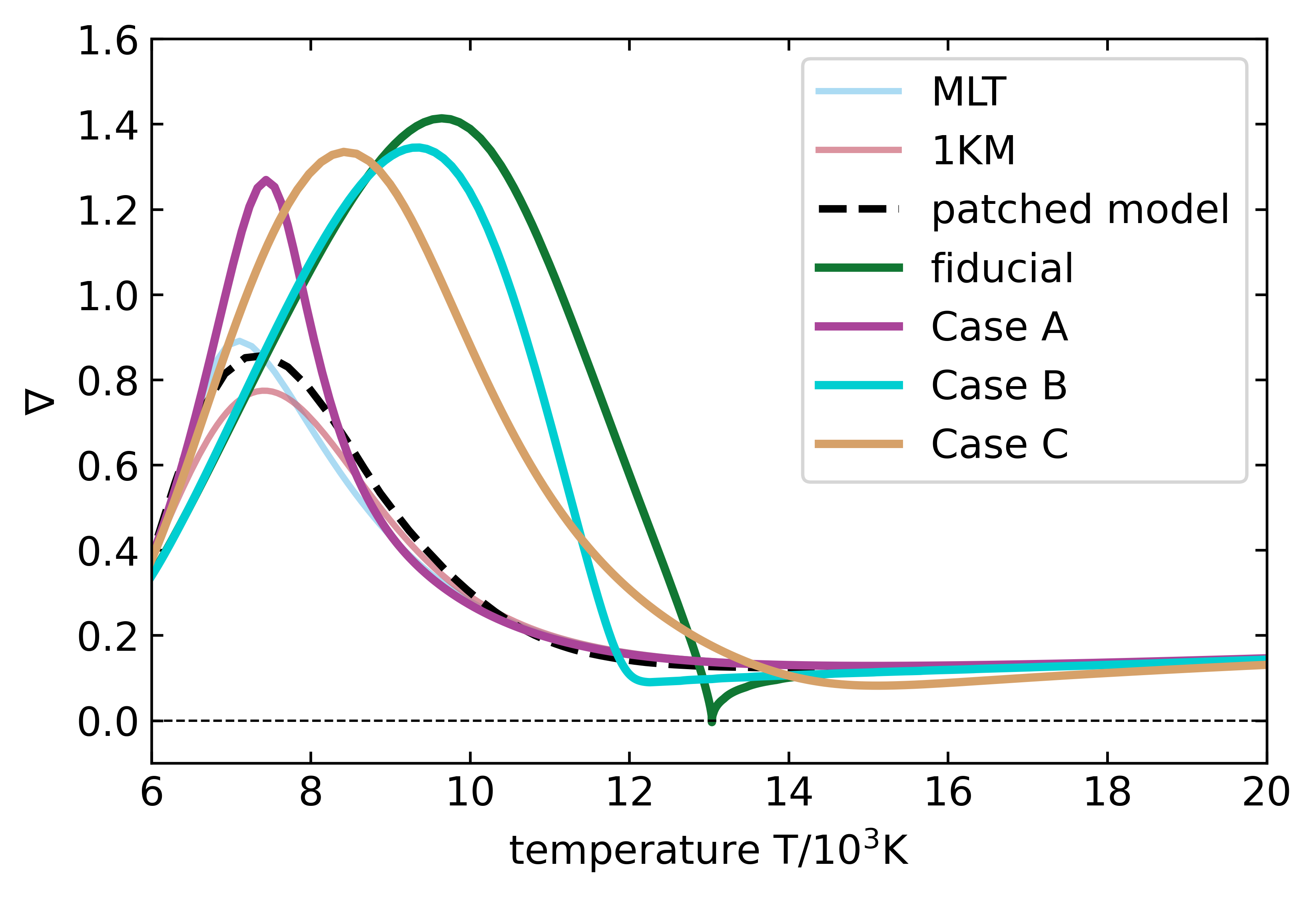}
    \caption{Temperature gradient against temperature of the uppermost layers. The temperature range shown corresponds to the uppermost 0.0034~\Rsun. The thin light blue and light pink lines show the profiles for \mlt\ and \ssmoneeq, respectively. The black, dashed line shows the profile of a solar model using an averaged 3D atmosphere as outer boundary condition \citep[``patched model'', ][]{Jorgensen2019}. The green, purple, turquoise, and orange lines show the profiles for the fully non-local \threeeq, and for case A, case B and case C, respectively (see Table~\ref{tab:cases}).}
    \label{fig:frankenstein-T-gradT}
\end{figure}

\begin{figure}
    \centering
    \includegraphics[width=1.0\linewidth]{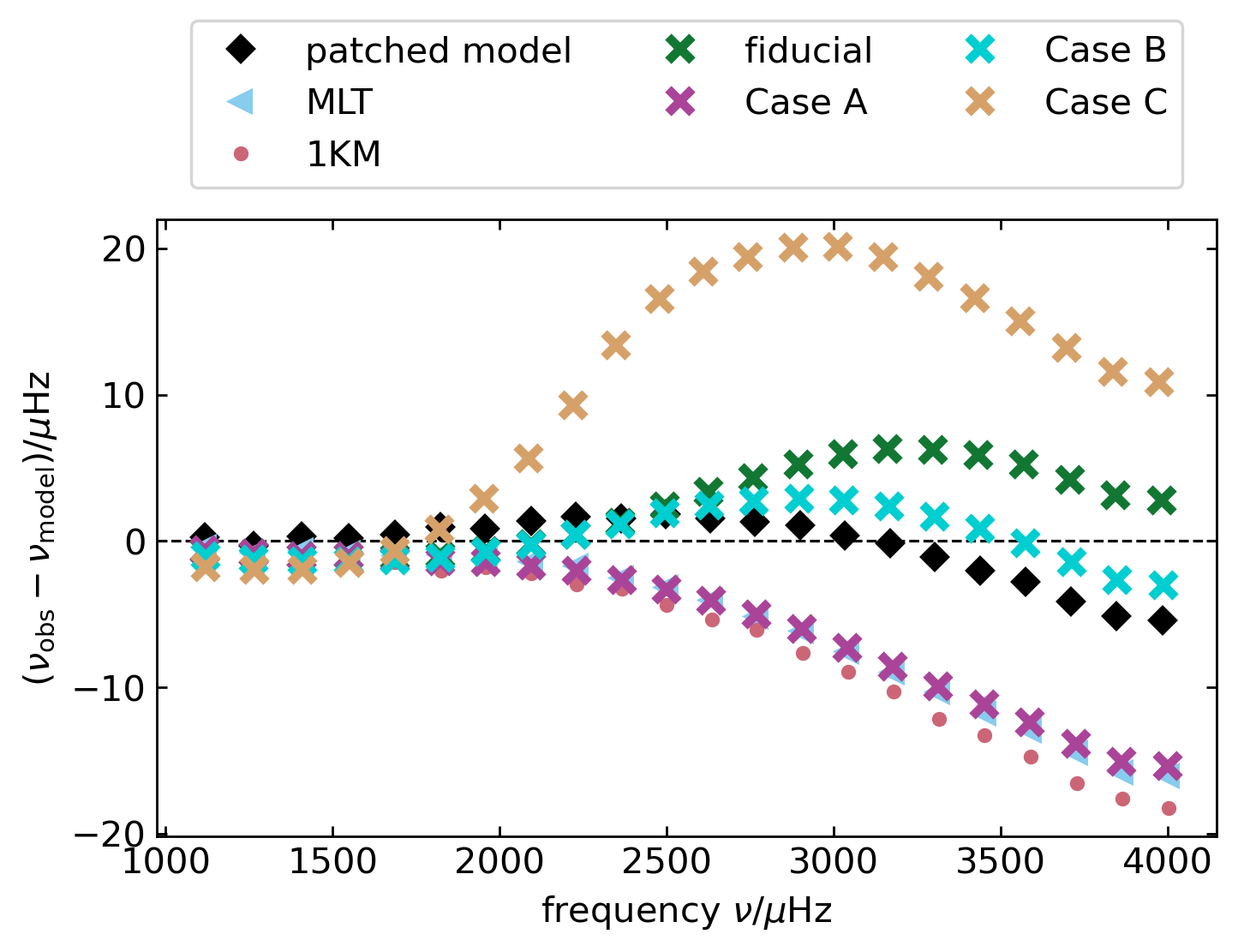}
    \caption{The surface effect of the solar models with different closure relations for the \threeeq\ in the outer regions (fully non-local: green; Case A: purple; Case B: turquoise; Case C: orange; see Table~\ref{tab:cases}). The result for \mlt\ (light blue triangles), \ssmoneeq\ (light pink dots), and the patched model \citep[black, ][]{Jorgensen2019} are shown for comparison.}
    \label{fig:frankenstein-surfaceeffect}
\end{figure}

\begin{figure}
    \centering
    \includegraphics[width=1.0\linewidth]{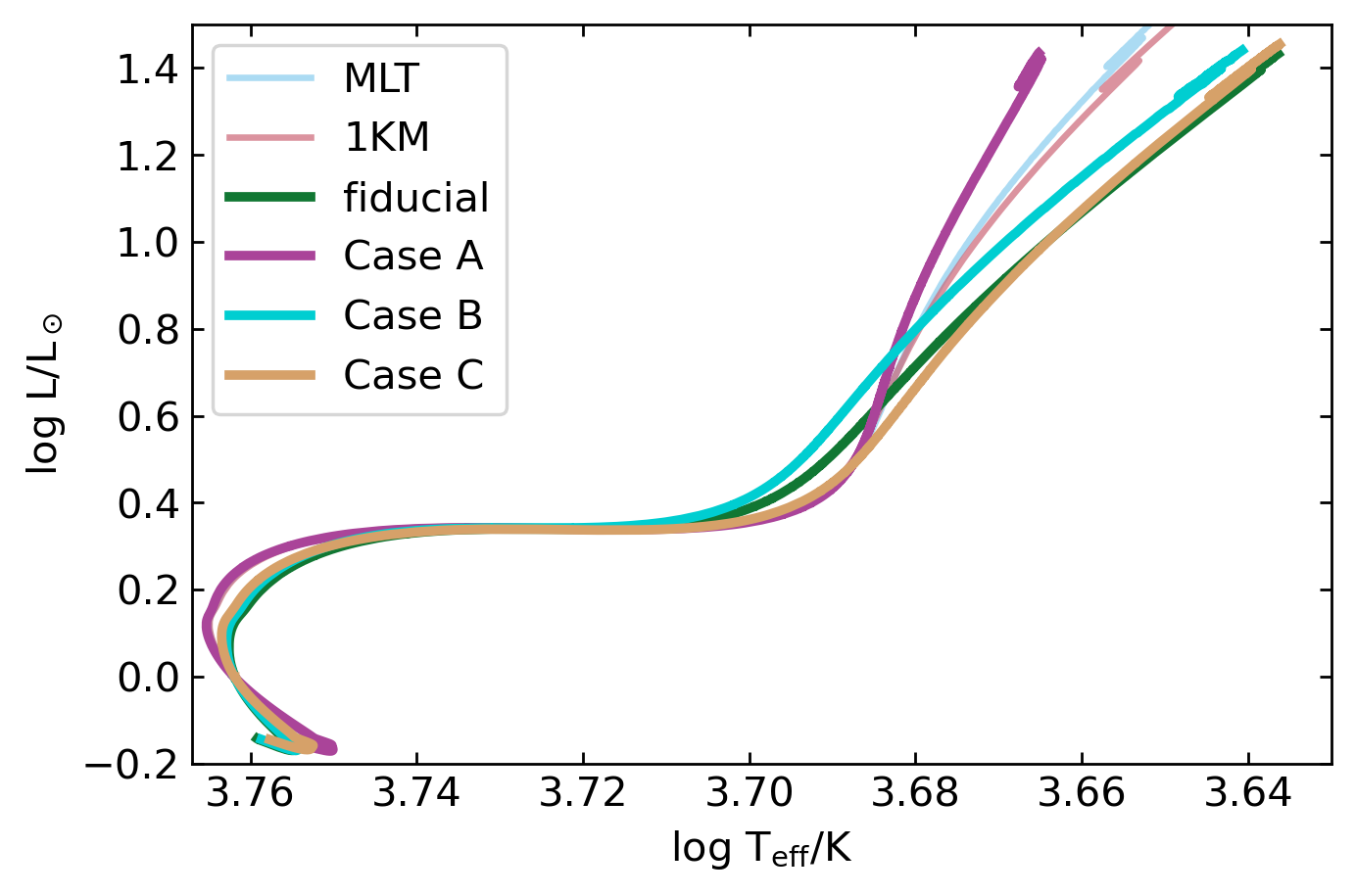}
    \caption{The HRD showing the evolution of the solar models. The models using MLT and the \oneeq\ are shown in thin light blue and light pink lines for comparison. The different treatment of the outer layers for the models with \threeeq\ result in higher (Case A: purple) and lower (fully non-local: green; Case B: turquoise; Case C: orange) effective temperatures on the RGB compared to MLT and \oneeq.}
    \label{fig:hrd}
\end{figure}

The profile of the temperature gradient of case A shows the most direct similarity to \mlt\ and \ssmoneeq\ (see Fig.~\ref{fig:frankenstein-T-gradT}). Except for a larger temperature gradient ($\nabla=1.27$), and a slight shift of the peak of the SAL to $7.4\cdot10^{3}$~K, it agrees with the profiles of \mlt\ ($\nabla=0.89$ at $7.1\cdot10^{3}$~K) and \ssmoneeq\ ($\nabla=0.77$ at $7.6\cdot10^{3}$~K). Directly below the SAL follows a weakly superadiabatic region (mean $\nabla-\nabla_\mathrm{ad}\sim1.8 \cdot 10^{-3}$). 
In the regions where the fully non-local \threeeq\ is applied, the behaviour of the temperature gradient is like in the other models. The similarity of MLT and case A is also visible in the surface effect (Fig.~\ref{fig:frankenstein-surfaceeffect}). The evolution of case A in the HRD is the same as for MLT and \oneeq\ until it reaches the lower RGB (see Fig.~\ref{fig:hrd}). Thereafter, the effective temperature is getting continuously higher compared to the models using MLT or \oneeq.

Both, case B and case C result in a subadiabatic region below the SAL (Fig.~\ref{fig:frankenstein-T-gradT}). The SAL is broader for case C ($\Delta T=7.9\cdot10^{3}$~K), the peak of this region is at $8.6\cdot10^{3}$~K, which is between the peak of the \mlt\ and the fiducial model. The change from a super- to a subadiabatic temperature gradient is very gradual. This results in a surface effect which is even more positive ($\nu_\mathrm{obs}>\nu_\mathrm{model}$, Fig.~\ref{fig:frankenstein-surfaceeffect}) than for the fiducial model. Case B has a more narrow SAL ($\Delta T=5.8\cdot10^{3}$~K), the change to the subadiabatic temperature gradient is less gradual. The peak of the SAL ($\nabla=1.3$ at $9.5\cdot10^{3}$~K) is closer to what was obtained with the fully non-local \threeeq. This profile results in frequencies which decrease the difference between observed and theoretical frequencies compared to the fully non-local case. This is also the model for which the surface effect is closest to what was obtained with the patched solar model, although the structure of the SAL is different. This again shows that it is not obvious to draw direct conclusions about the surface effect from the temperature stratification in the outer layers alone. In the HRD, both, case C and case B, are at lower effective temperatures along the RGB. Case C starts the RGB at cooler temperatures but follows the fully non-local \threeeq\ closely afterwards. The effective temperature on the RGB for case B is between the one obtained when using MLT or \oneeq\ and the fully non-local \threeeq.

The shift in effective temperature between the different cases is the result of different convective fluxes in the models. 
The higher convective flux of case A causes the radius to be smaller, which means the temperature is higher at a given luminosity. The lower effective temperatures of the fiducial model, case C, and case B are due to their lower convective flux. The thin layer at which the different closure relations are used affects the bulk of the convective zone enough to result in the shift in effective temperature seen on the RGB, although the temperature stratification at the inner boundary is not affected (see Sect.~\ref{sec:disc:literatur} for a discussion about observational uncertainties).

Since none of these cases show a negative temperature gradient, it is clear that the unrealistic modelling of the outer layers is caused by the closure relations of the $\Phi$- and/or $\Pi$-equation. The coupling of the equations makes it difficult to see exactly which of the two is the main contributor to this behaviour, however, improving those terms can be an ansatz to further develop the 3-equation Kuhfuss model \citep{Kupka2026}.

\section{Discussion}\label{sec:discussion}

In Sect.~\ref{sec:disc:convflux}, we discuss the main difference between the \oneeq\ and \threeeq\ and the effect of it, that is, the application of the downgradient approximation for the convective flux in \oneeq\ which in \threeeq\ got replaced by solving Eqs.~\eqref{eq:pi} and ~\eqref{eq:phi}. The results are compared to the literature in Sect.~\ref{sec:disc:literatur}. 

\subsection{Differences in the Convection Models}\label{sec:disc:convflux}

Apart from the inclusion of non-local effects, which are also included in \oneeq, the most significant difference between MLT and \oneeq\ compared to the \threeeq\ is the modelling of the convective flux, which influences and explains the profile of $\nabla$ in the \ssmthreeeq.
MLT and \oneeq\ both assume that the convective flux is proportional to the entropy gradient, that means, 
\begin{equation}
    F_\mathrm{conv} \propto - \frac{\partial s}{\partial r} \propto \nabla - \nabla_\mathrm{ad} \, ,
\end{equation}
which ties the sign of $F_\mathrm{conv}$ to the superadiabaticity of the layer. For \threeeq, this downgradient approximation is not applied (see Eq.~\ref{eq:pi}) and the convective flux has contributions from the superadiabatic gradient and from the entropy fluctuations. This, together with the inclusion of the non-local effects of the entropy fluctuations \citep{Deardorff1966}, makes it possible to have a Deardorff layer: a layer with $\nabla-\nabla_\mathrm{ad}<0$ and $F_\mathrm{conv}>0$  \citep{Deardorff1966, Chan1992, Tremblay2015, Kapyla2017, Andrassy2024, Kapyla2025}. It also weakens the coupling between the chemical mixing from the modifications of the thermal structure. Thus, the mixed region extends into layers with a close-to-radiative temperature gradient, another feature which is not obtained with the \oneeq\ and similar convection models.

Comparing Eq.~\eqref{eq:pi}, the equation describing the convective flux, with Eq.~(14) from the derivation by \citet{Brandenburg2016}, it becomes clear that he arrives at a very similar model as \threeeq\ from a slightly different ansatz. With this ansatz, \citet{Brandenburg2016} finds convection zones with extended Deardorff layers, with long narrow plumes reaching from close to the surface far into regions with a subadiabatic temperature gradient. Those plumes are highly non-local and are driven by the surface cooling. This effect is called ``entropy rain''. In 3D simulations, Deardorff layers often only extend over a limited range of the convective region \citep{Kapyla2017}, similar to what \threeeq\ predicts for the Sun. However, \citet{Kapyla2025} tested if an extended entropy rain, as discussed by \citet{Brandenburg2016}, can be realised in 3D simulations. Instead of assuming uniform cooling, \citet{Kapyla2025} applied non-uniform cooling patches to the upper boundary of the 3D simulation, taking the non-uniform surface temperature of the Sun as inspiration. He found that with this set-up, the bulk of the convective zone is subadiabatic with fast, narrow downflows. From observational data, \citet{Bekki2024} found evidence that the bulk of the solar convective envelope is less superadiabatic than usually assumed, or even subadiabatic. These studies indicate that subadiabatic convection may be present in the solar envelope. This would imply that MLT does not accurately represent the temperature stratification due to its underlying assumptions. The \threeeq\ constitutes an important step towards a more realistic representation of such subadiabatic convection in 1D stellar evolution codes.

\subsection{Comparison to the Literature}\label{sec:disc:literatur}

The determination of $R_\mathrm{cz}$ involves comparing and minimizing the difference of the sound speed between the Sun and a reference model \citep{Basu1997b}. The reference models used by \citet{Basu1997b} to determine $R_\mathrm{cz}$ either did not include overshooting or did include it in the sense of adiabatic overshooting, that means the temperature gradient in the CBM region is assumed to be close-to-adiabatic with a sharp transition to $\nabla_\mathrm{rad}$ at the boundary. With these reference models, they found that the change in the temperature gradient happens at a radius of $0.713\pm0.001$~\Rsun\ \citep{Basu1997b}. It is unclear if this result is directly applicable to the \ssmthreeeq, because of the much smoother change in $\nabla$ compared to the reference models used to derive this value.
Repeating the measurement of $R_\mathrm{cz}$ with models with a smooth transition in the temperature gradient, such as the \ssmthreeeq, is needed to assess what effect the profile of $\nabla$ of the stellar model has on the measurement of $R_\mathrm{cz}$.

The measurement of $R_\mathrm{cz}$ is further complicated by composition gradients at the base of the convective zone \citep{Basu1994}.
\citet{Basu1997b} used different models to determine what composition profile fits the signatures from helioseismology best. They found that only models with a smooth composition profile are consistent with the observations. The additional feature in the derivative of the sound speed of \ssmthreeeq\ indicates that \threeeq\ predicts a composition profile which is too sharp. A more gradual damping of the TKE or additional effects such as shear flows due to rotation \citep{Richard1996, Basu1997a} may help to smooth out the composition profile in the transition region from convective to radiative. 

\citet{Christensen-Dalsgaard2011} parametrized the temperature stratification below the convective region to be able to vary how smoothly the temperature gradient changes from close to $\nabla_\mathrm{ad}$ to $\nabla_\mathrm{rad}$. The temperature stratification which \citet{Christensen-Dalsgaard2011} found to agree with the helioseismic data best is in qualitative agreement with the profile predicted by \threeeq. 
A smooth change from $\nabla_\mathrm{ad}$ to $\nabla_\mathrm{rad}$ was also found in the 3D hydrodynamical simulations from \citet{Kapyla2017}. 
\citet{Xiong1989} developed a TCM similar to the \threeeq, which was extended to include anisotropic effects by \citet{Deng2006}. \citet{Xiong2001a} used this convection model to calculate an envelope model for the Sun, which was used as reference model for helioseismic inversions by \citet{Zhang2012b}. For an overview of this convection theory and the tests performed, see \citet{Xiong2021}, and references therein.
They found that the transition of $\nabla$ from close-to-adiabatic to radiative is smooth and that $\nabla$ is already subadiabatic within the formally unstable layer \citep{Xiong2001a}. They also found that this improves the sound speed profile \citep{Zhang2012b}.
Similar results were found by \citet{Zhang2012c}. They used the TCM developed by \citet{Li2007} to calculate an evolutionary model of the Sun. Again, using a TCM improved the sound speed profile and lead to a smoother transition of the temperature gradient in the CBM region.
The \threeeq\ is in qualitative agreement with these other convection models regarding the smoother transition in the temperature gradient at the lower boundary and the resulting improvement in the sound speed profile. However, these other works do not predict a subadiabatic region close to the SAL, as is the case when using \threeeq.

\citet{Rempel2004} approached the problem of the CBM at the base of the solar convection zone with a semianalytical convection model, considering individual plumes reaching from the top of the convective zone to the bottom. With this model, they also found an extended Deardorff layer and a smoother transition from a close-to-adiabatic to a radiative temperature gradient at the lower boundary of the convective zone. The smoothness of the transition of the temperature gradient depends mainly on the free parameter which controls the ratio between the dimensionless total energy flux and the filling factor of the downward plumes at the base of the convective envelope. 
They argue that non-local extensions of MLT result in adiabatic overshoot not because of crude approximations but because of the assumption of a large filling factor. 
The difference between the model of \citet{Xiong2001a} and the one by \citet{Rempel2004} is concluded in the latter to stem from the dominant breaking process in the CBM region, which is buoyancy breaking in \citet{Rempel2004} as opposed to turbulent dissipation in \citet{Xiong2001a}. In the \threeeq, the dominant breaking process in the CBM region is also turbulent dissipation. In contrast to the model by \citet{Xiong2001a}, however, \threeeq\ takes dissipation due to buoyancy waves into account \citep{Kupka2022}. 
In conclusion, the different convection models obtain the same result of a Deardorff layer and a smooth transition to the radiative temperature gradient despite considering different physical processes.

The effective temperature on the RGB is cooler when \threeeq\ is used compared to \oneeq\ or MLT. This shift does not immediately rule out the \threeeq because stellar models with a solar calibrated $\alpha_\mathrm{MLT}$ were found to be in disagreement with some observations \citep{Bonaca2012, Creevey2015, Joyce2018a, Joyce2018b, Viani2018}. In particular, \citet{Tayar2017} found that stellar models with a solar calibrated $\alpha_\mathrm{MLT}$ are in tension with the effective temperatures of red giants, although this study is challenged by \citet{Salaris2018} and \citet{Choi2018}. Detailed studies are needed before we can conclude whether the change in effective temperature on the RGB for solar models using \threeeq\ is desirable or not.

The closure relations for the TOMs of the $\Pi$- and $\Phi$-equation have a strong effect on the SAL, and the subadiabatic layer below it. Therefore, they also influence the effective temperature on the RGB. Improving the closure relations is needed to remove the artifact of the temperature inversion in these layers when using the \threeeq\ with the non-local closure relations for all equations. Work developing new closure relations for \threeeq\ is currently under way \citep{Kupka2026}.

\section{Conclusion and Summary}\label{sec:conclusion+summary}

We calculated a solar model using the non-local 3-equation Kuhfuss turbulent convection theory \citep[\threeeq, ][]{Kuhfuss1987, Kupka2022, Ahlborn2022}. Apart from the non-locality, the major difference between \threeeq\ and classical MLT is the inclusion of the second order entropy fluctuations contributing to the convective flux. 

The \threeeq\ improves the modelling of the convective boundary mixing region. It reproduces the temperature stratification at the inner boundary of the convective zone in qualitative agreement to the optimal stratification inferred by \citet{Christensen-Dalsgaard2011} from helioseismic measurements. \threeeq\ also predicts a Deardorff layer, which is especially interesting as first observations suggest that the solar convective envelope is, in some regions, less superadiabatic than previously assumed, or maybe even subadiabatic \citep{Bekki2024}.
The inclusion of the second order entropy fluctuations is crucial to obtain a Deardorff layer and a smooth transition of the temperature gradient in the convective boundary mixing region. 
Furthermore, the agreement of the sound speed profile between helioseismic measurements and the stellar model is significantly improved.

While significantly improving the lower boundary of the convective envelope, problematic features arise at the upper boundary. Right below the superadiabatic layer, the temperature gradient becomes strongly subadiabatic and even negative, before it becomes superadiabatic again. This is a feature which is not observed in 3D simulations, and is not described in the literature covering other 1D descriptions of non-local convection.
We investigated if these features are connected to a specific choice of parameters, but we could not confirm a clear correlation between one specific free parameter and the features described above. Instead, we showed that different combinations produce solar models with similar features but different details. Due to the interplay between the free parameters, it is not possible to calibrate all of them with the limited observations we have. To further constrain their values, a comparison with 3D simulations is needed. 

With the goal to better understand which term is responsible for the unrealistic behaviour just below the superadiabatic layer, we calculated models using the local closure relations (Eq.~\ref{eq:local-closure}) for the $\Pi$- and $\Phi$-equation in the outer $\approx$0.021~\Rsun. The inner boundary of the convective region is not affected significantly. In the outer layers, this results in a temperature stratification which is in better agreement with MLT and 3D simulations because the temperature gradient does not become negative any more. However, there is still a subadiabatic region between the superadiabatic peak region and a second superadiabatic region, which is likely unphysical and not observed in the 3D simulations.
Since these equations are coupled and all versions result in a non-negative temperature gradient, it is not obvious which of the two terms is the main contributor. What becomes clear is that improving the closure relation of the $\Pi$- and $\Phi$-equation is a promising ansatz to further develop this model \citep[see][]{Kupka2026}. 

Thus, the \threeeq\ can serve as a starting point to develop a new convection theory. While still showing some limitations, the \threeeq\ is clearly a step forward to improve the modelling of convection in 1D stellar evolution codes.

\begin{acknowledgements}
We thank Sarbani Basu for productive discussions of the helioseismic observables. We thank Yago Herrera for providing the opacity tables with the updated solar composition. We thank Andreas J{\o}rgensen for providing the data of the patched solar model. FK is grateful for the hospitality provided by the Wolfgang Pauli Institute, Vienna, and acknowledges support by the Faculty of Mathematics at the University of Vienna through offering him a Senior Research Fellow status --- the Austrian Science Fund (FWF) has supported this research through grants P~33140-N  and P~35485-N. FA acknowledges funding from the European Research Council under the European Community’s Horizon 2020 Framework/ERC grant agreement no 101000296 (DipolarSounds) and thanks the Klaus Tschira foundation for their support.
\end{acknowledgements}

\bibliographystyle{aa}
\bibliography{bib}

\begin{appendix}

\section{Solar Model with the Abundances from \citet{Asplund2009}}\label{ap:asplund-abundances}
    
The composition of the Sun is an open question.  
\citet{Magg2022} published abundances of the Sun based on averaged 3D simulations. They found that the surface metal-to-hydrogen ratio of the Sun is $Z_\odot/X_\odot=0.0225$. This is close to the value derived by \citet{Grevesse1998} who used 1D atmosphere models to obtain this value. On the other hand, \citet{Asplund2009, Asplund2021} used 3D simulations and determined the metal-to-hydrogen ratio to be lower compared to \citet{Magg2022} (\citealp{Asplund2009}: $Z_\odot/X_\odot=0.0181$; \citealp{Asplund2021}: $Z_\odot/X_\odot=0.0187$).

To address this conflict, \citet{Buldgen2023} used helioseismic inversions to derive a solar metal mass fraction independent of spectroscopic models. They found that a low metallicity, more similar to \citet{Asplund2009, Asplund2021}, is favoured. \citet{Buldgen2019b} argue that the solar modelling problem is not only a question about the solar abundances but is also affected by ingredients such as the equation of state or the opacities \citep{Buldgen2025a}. This can potentially explain the larger disagreement in the sound speed profile when using the lower metallicities.

Due to this unsettled question, in this section, we present our results for the abundances from \citet{Asplund2009}. All parameters except for $\alpha_\Pi$ have the same values as the fiducial model of Sect.~\ref{sec:solcalib} with the abundances by \citet{Magg2022}. The composition of the opacity tables is again consistent with the abundances. The characteristics of the solar calibrated model using the abundances from \citet{Asplund2009} are included in Table~\ref{tab:test-calib} (labelled ``Asplund'').

\begin{figure}[htb!]
    \centering
    \includegraphics[width=1.0\linewidth]{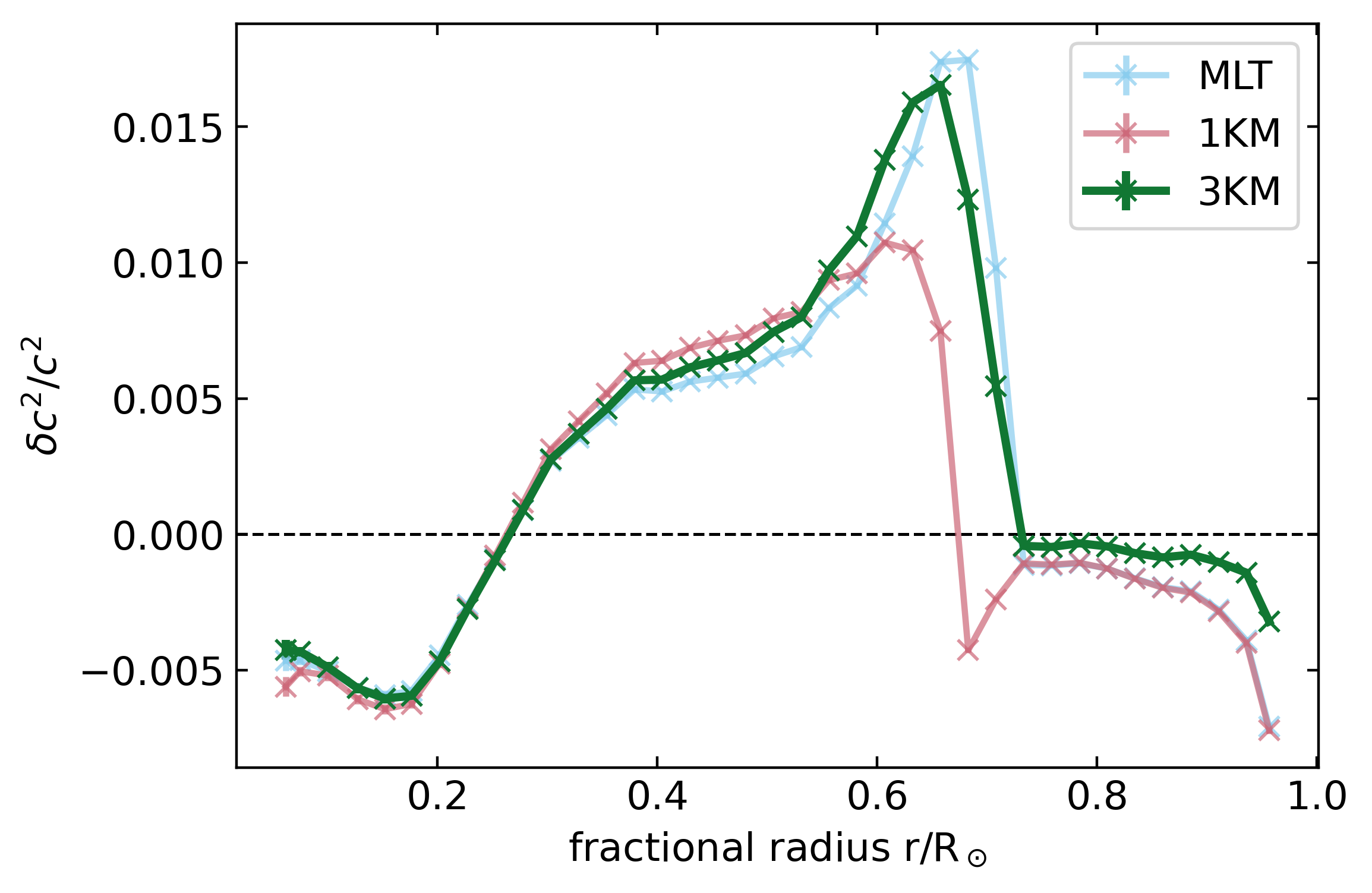}
    \caption{The relative difference of the squared solar sound speed $c_\mathrm{helio}$ and the sound speed obtained from models using the \citet{Asplund2009}-abundances $c_\mathrm{model}$: $\delta c^2/c^2 = (c^2_\mathrm{helio}-c^2_\mathrm{model})/c^2_\mathrm{helio}$. The solar models using MLT (light blue), the \oneeq\ (light pink), and the \threeeq\ (green) are shown.}
    \label{fig:cs-asplund}
\end{figure}

As for MLT, the \citet{Asplund2009}-abundances bring the boundary of the convective region to larger radii, which improves the agreement with the measurement to 2.5$\sigma$. The He abundance in the envelope of the solar model with \threeeq\ and the \citet{Asplund2009}-abundances is $Y_\mathrm{cz}=0.2379$, which is within 3.1$\sigma$ of the measurement. 
Like in the model using \citet{Magg2022}-abundances, a change in the sound speed derivative is visible at the radius where the TKE becomes zero. The SAL of the model using \citet{Asplund2009}-abundances has generally the same properties as the SAL from the model with \citet{Magg2022}-abundances. Figure~\ref{fig:cs-asplund} shows the sound speed profile for the models with \citet{Asplund2009}-abundances. As for MLT and \oneeq, the deviation around the boundary of the convective region is larger compared to the models using the higher metal abundances from \citet{Magg2022}. Interestingly, the \threeeq\ does not cause an improvement in this region as it does for the models with \citet{Magg2022}-abundances. Instead, the result with the \threeeq\ is very similar to the result with MLT.

\section{Test of the Effects of Calibration}\label{ap:test-calibration}

The models with different combinations of the free parameters (Sect.~\ref{sec:varyparms}) are calibrated to a lower accuracy than the fiducial model from Sect.~\ref{sec:solcalib}. This is because of the time-consuming calibration process, since only small adjustments can be made in each calibration step due to the difficult convergence of the model.
To investigate which effects are caused by a lower accuracy of the calibration, and which are actually caused by the different values for the free parameters of the convection theory, we calculated solar models with a lower accuracy than the fiducial model and compared them. The accuracy of these models and the resulting values of $\alpha_\Pi$ are given in Table~\ref{tab:test-calib}.
    
\begin{table*}[htb!]
\caption{Testing the effect of using the abundances from \citet{Asplund2009} (Appendix~\ref{ap:asplund-abundances}), and of models with lower accuracy (Appendix~\ref{ap:test-calibration})}  
\label{tab:test-calib}
    \centering
    \begin{tabular}{c|c|ccc|cc}
    \hline \hline
       Name & $\alpha_\Pi$  & $\delta R$/R$_\odot$ & $\delta L$/L$_\odot$ & $\delta (Z/X)$/(Z$_\odot$/X$_\odot$) & $Y_\mathrm{cz}$  & $R_\mathrm{cz}$  \\
         &    & [$10^{-4}$] & [$10^{-4}$] & [$10^{-4}$] &    &  [\Rsun]  \\
    \hline  
    asplund            &  2.22    &  1.1   & -1.5  & -20   &  0.2379  &  0.7155   \\ \hline
    $\alpha_\Pi=2.102$ &  2.102   & -20  &  4.4  &  7.8  &  0.2456  & 0.7048  \\
    fiducial           &  2.155   &  0.66  & -0.89 &  3.3  &  0.2455  & 0.7067  \\
    $\alpha_\Pi=2.2$   &  2.2     &  17  & -6.1  & -0.88 &  0.2455  & 0.7084  \\
    \hline
    \multicolumn{7}{l}{\small All other free parameters of \threeeq\ have the same values as used for the fiducial model (see Table~\ref{tab:vary-parms}).} \\ 
    \end{tabular}
\end{table*}

\begin{figure}[htb!]
    \centering
    \includegraphics[width=1\linewidth]{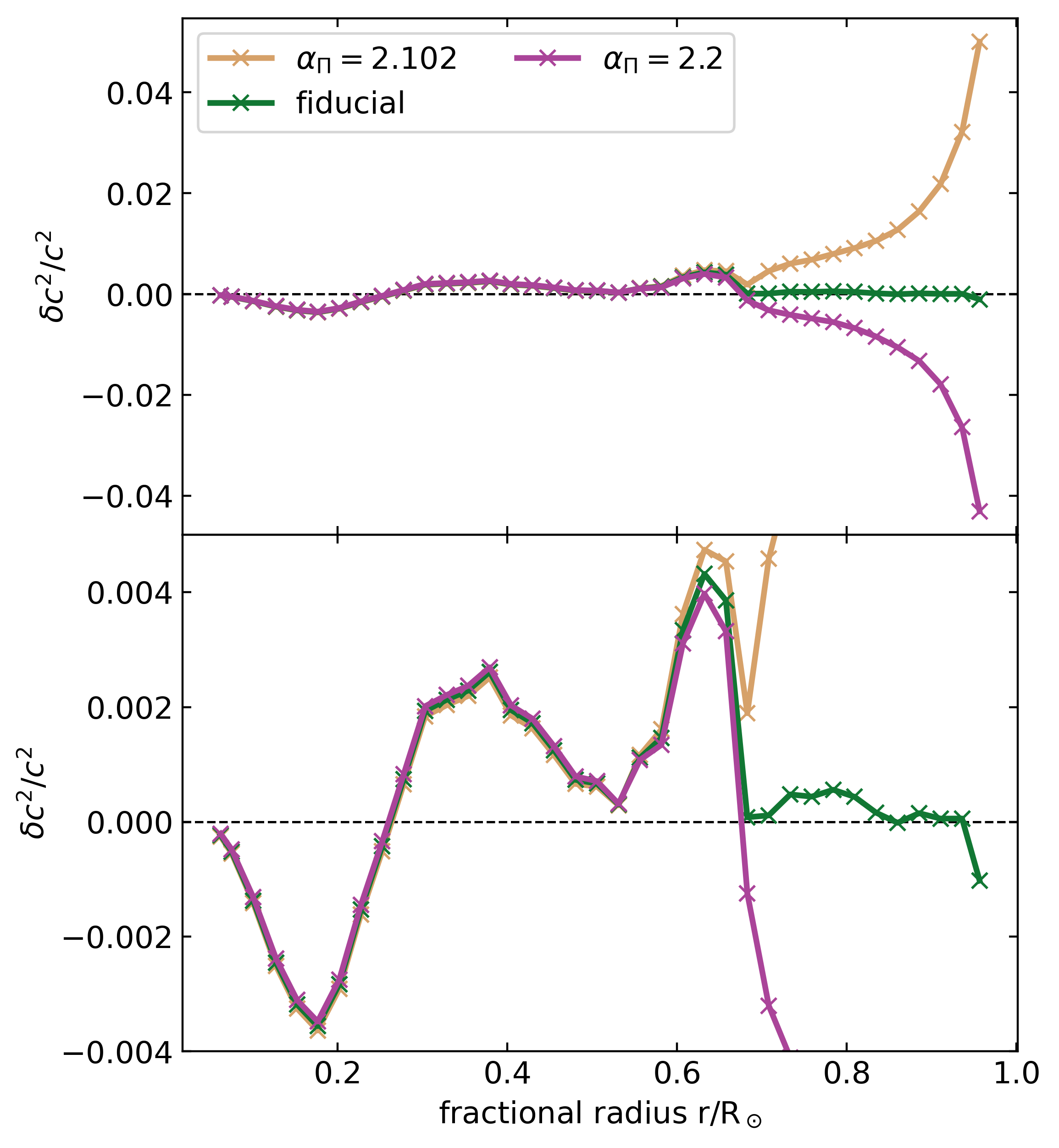}
    \caption{The relative difference of the squared solar sound speed $c_\mathrm{helio}$ and the sound speed obtained from models with varying accuracy of the calibration $c_\mathrm{model}$: $\delta c^2/c^2 = (c^2_\mathrm{helio}-c^2_\mathrm{model})/c^2_\mathrm{helio}$. The upper panel shows the complete range, while the lower panel zooms in to show more details.}
    \label{fig:test-calib}
\end{figure}

Figure~\ref{fig:test-calib} compares the sound speed profiles of the three models given in Table~\ref{tab:test-calib}. The sound speed of the outer layers is heavily influenced by the radius of the model. Because the models were calibrated to a lower accuracy, the difference in the sound speed between the observation and the models with $\alpha_\Pi=2.2$ and $\alpha_\Pi=2.102$ reaches absolute values of $\approx0.05$ in the outer layers. Compared to the fiducial model, this increases the absolute difference in $\delta c^2/c^2$ by 0.05, and 0.04, for the model with $\alpha_\Pi=2.102$ and $\alpha_\Pi=2.2$, respectively. However, the feature close to the boundary of the convective envelope is less affected, as can be seen in the lower panel of Fig.~\ref{fig:test-calib}. In this region, compared to the fiducial model, $\delta c^2/c^2$ is decreased by $3.4\cdot 10^{-4}$ for the model with $\alpha_\Pi=2.2$, and increased by $4.2\cdot 10^{-4}$ for the model using $\alpha_\Pi=2.102$.

The difference in $R_\mathrm{cz}$ between the models reflects the general difference in the exact radius of the models. 
The absolute extent of the convective envelope, $\Delta R_\mathrm{cz} = R_\mathrm{cz} - R_\star$, is barely changed. Compared to the fiducial model, it is increased by $2\cdot 10^{-5}$~\Rsun\ for the model with $\alpha_\Pi=2.102$, and decreased by $7 \cdot 10^{-5}$~\Rsun\ for the model with $\alpha_\Pi=2.2$. Also, the He content of the envelope is not affected (see Table~\ref{tab:test-calib}).

Therefore, one can conclude that a lower accuracy of a calibrated solar model will mainly affect the outer parts of the sound speed profile, while the effect on the inner boundary of the convective envelope is minor.

\section{Settings for GYRE}

We used the following settings for the control file for GYRE. If not stated otherwise, we used the default settings. We used the outer boundary conditions and dependent variable set following ADIPLS \citep{Christensen-Dalsgaard2008}. The fourth-order Gauss-Legendre Magnus difference equation scheme was used. For the frequency scan parameters, we used \texttt{freq\_min = 1000}, in units of $\mu$Hz, \texttt{freq\_max = 1} in units of the acoustic cut-off frequency, and \texttt{n\_freq = 100}. Furthermore, we used \texttt{w\_ctr = 10}, \texttt{w\_osc = 10}, \texttt{w\_exp = 4}.
    
\end{appendix}
\end{document}